\documentclass{article}
\usepackage{amsmath}
\usepackage{amssymb}
\usepackage{hyperref}
\usepackage{graphicx}
\newcommand{\HALF}{\frac{1}{2}}
\renewcommand{\vec}[1]{\mathbf{#1}}
\newcommand{\hvec}[1]{\hat{\mathbf{#1}}}
\newcommand{\pd}[2]{\frac{\partial #1}{\partial #2} }
\newcommand{\DS}{\displaystyle}

\newcommand{\Hyd}{{\textsc{\tiny H}}}
\newcommand{\Hel}{{\textsc{\tiny He}}}
\newcommand{\X}{{\textsc{\tiny X}}}

\usepackage{authblk} % Package for managing authors and affiliations
\usepackage{setspace} % Package to adjust line spacing

\title{\textbf{Simulations of Protostar-Driven Photoionization in Herbig-Haro Jets}}

\author[1,2]{Z. Ahmane}
\author[2]{A. Mignone}
\author[2]{C. Zanni}
\author[2]{S. Massaglia}
\author[1]{A. Bouldjderi}

\affil[1]{Hadj Lakhdar Batna 1 University, Algeria}
\affil[2]{Università degli Studi di Torino, Italy}

\affil[ ]{\texttt{zoubida.ahmane@univ-batna.dz} (corresponding author)}

\date{} 

\begin{document}
\maketitle

% === Advisor suggestion 2: Note about the published version ===
\vspace{0.5em}
\noindent This arXiv version is a revised and updated version of the peer-reviewed paper published in \textit{Astrophysics and Space Science}, 2020, Vol. 365, 94.
\section*{Abstract}
Recent studies showed that observations of line emission from shocks in YSO jets require a substantial amount of ionization of the pre-shock matter.  
Photoionization from X-rays emitted close to the central source may be responsible for the initial ionization fraction.  
The aim of our work is to study the effect of X-ray photoionization, coming  from the vicinity of the central star, on the ionization fraction inside the jet that can be advected at large distances. 
For this purpose we have performed  axisymmetric MHD jet launching simulations including photoionization and optically thin losses using PLUTO.  
For typical X-ray luminosities in classical T-Tauri stars, we see that the photoionization is responsible for ionizing  to 10 \% -20 \% of the jet close to the star.\\

\textbf{keywords} YSO- jets , X-rays,  Hydrodynamics, Methods: Numerical, Magnetohydrodynamics: MHD, Astrophysics

\section{Introduction}
A protostellar jet is a highly supersonic, magnetised and collimated outflow that heralds the birth of a star while it remains deeply embedded and invisible inside its molecular cocoon.

In particular, Herbig-Haro (HH) jets such as HH30, DG Tau and RW Aur \cite{BE99, LA97, LA00, BA02, ME09} are observed in star-forming regions around T Tauri stars. They are characterised by highly collimated ejection of material that originates directly from the innermost regions surrounding the protostar \cite{RE86, MU85, LA85, MA14}. The associated line emission arises from shocked gas that has been heated and compressed \cite{SC75, DO78} by nonlinear magnetohydrodynamic (MHD) effects. These include shear-layer instabilities \cite{MA92, HA97, MI00} and/or time variability in the accretion/ejection process that steepens into internal shocks \cite{RA95, DE08}.

Modelling these magnetised supersonic shocked outflows, whether theoretically or numerically, requires the inclusion of non-equilibrium cooling and heating processes. Such modelling is essential for interpreting jet phenomenology and for constraining the main physical parameters of the jets. This explains the extensive effort devoted by many authors to simulating jet dynamics together with radiative effects \cite{RO97, SH02, MO06, TE08, TE12, TE14, RU14, GA17}.

The basic properties of protostellar outflows have been known for decades. The outflow phase lasts for at least $10^{5}$ yr (e.g.\ in Class I stars). These flows extend over lengths of $\geq 10^{3}-10^{4}$ AU \cite{BA07}, reach velocities of $v_{\mathrm{jet}} = 100-800$ km s$^{-1}$ that increase with central stellar mass \cite{EM92, EA94}, and are very well collimated with typical radii of $\sim 50$ AU \cite{RA96, DO00}. They are powered by accretion onto the central protostar \cite{CA90, HA95, LI99}.

The jet acceleration mechanisms proposed in the literature are: (i) the magneto-centrifugal launch of the so-called ``Disk-Winds'' \cite{BL82}, (ii) thermal pressure at the base of the jets that produces ``stellar winds'' \cite{SA94, SA02}, and (iii) the star-disk magnetospheric interactions that produce ``X-winds'' \cite{SH94, CA08}. Both disk-wind and stellar-wind mechanisms can operate simultaneously, as discussed by \cite{MA08, MA09}.

Numerical simulations of HH jets must cover enormous spatial scales, from a few AU near the source up to tens of thousands of AU, because emission is observed on all scales and originates in post-shock regions. Line emission at these different scales provides valuable information on the physical parameters and, indirectly, on the jet morphology. Even with the most advanced Adaptive-Mesh-Refinement (AMR) techniques, performing a global, fully 3D MHD simulation that simultaneously resolves the jet launching region and the jet termination at sufficient resolution to capture internal shocks remains computationally impossible.

Therefore, large-scale simulations usually start at about 100 AU from the origin and inject the jet using reasonable but somewhat arbitrary initial and boundary conditions. The models then follow the propagation of the jet and its interaction with the ambient medium \cite{MO06, FA02, ST97, TE12, TE14}. These simulations can incorporate radiative cooling and heating and generate synthetic emission-line maps that allow detailed comparison with observations.

A complementary strategy is to simulate the jet launching and propagation self-consistently up to about 100 AU. This is achieved either by evolving the accretion disk and jet together in the computational domain \cite{UC85, CA02, ME06, ZA07, TZ09, TE12} or by imposing the disk as a fixed boundary condition \cite{OU97, FE09, FE06, MA12}. Axisymmetric simulations that treat both launching and large-scale propagation simultaneously have also been performed \cite{RA11}, although still within the ideal-MHD approximation.

The observed line emission is produced by the shocked gas immediately behind the shock front. Radiative shocks have been studied in 1D steady state by \cite{CO85, HA94}, who derived the post-shock evolution of key physical parameters (temperature, ionization fraction, electron density, etc.). The 1D time-dependent evolution of magnetised radiating shocks was investigated by \cite{MA05, TE09}. Two-dimensional numerical studies of radiating magnetised shocks in axisymmetric jets were carried out by \cite{TE12} using sinusoidal perturbations imposed at the jet base.

\cite{TE12} showed that an ionization fraction of $\sim 10-20\,\%$ is required to reproduce the observed brightness distributions in different emission lines. They proposed that this ionization is provided by high-energy photons from the embedded luminous protostar, an X-ray source with luminosity $L_{\X}= 10^{28} - 10^{32}$ erg s$^{-1}$ \cite{FA02, BA03, SK11}, and that the ionization persists over long distances along the jet due to the long recombination timescales.

The goal of the present paper is to test the validity of this hypothesis by performing simulations that treat the launching region and the large-scale propagation simultaneously, while consistently including MHD processes, X-ray photoionization heating, and radiative cooling. Given the large dynamical range involved, the simulations are performed in 2D using adaptive mesh refinement techniques.

The plan of the paper is the following: in Sect. 2 we discuss the model that includes cooling and photoionization due to X-rays, in Sect. 3 we outline the initial conditions and the computational scheme adopted; in Sect. 4 we examine the effect of X-rays on the ionization of the jet and the shock evolution, while in Sect. 5 we discuss the results and make comparisons with observations. Conclusions are drawn in Sect. 6.
%%%%%%%%%%%%%%%%%%%%%%%%%%%%%%%%%%%%%%%%%%%%%%%%%%%%%%%%%%%%%%%%%%%%%%%%%

\section{Model}

Our model consists of a time-dependent jet accelerated by an underlying accretion disk imposed as a boundary condition.
The equations of magnetohydrodynamics are solved using the PLUTO code \cite{MI07, Mig-PLUTO2012} by taking into account non-ideal effects due to radiative losses.

\subsection{Governing equations}

In what follows, the fluid density, velocity, magnetic field and thermal pressure will be denoted, respectively, with $\rho$, $\vec{v}$, $\vec{B}$ and $P$.
The gas pressure depends on the plasma density $\rho$, temperature $T$ and composition through the relation $P= \rho k_{B}T/(\mu m_{H})$, where $\mu$ is the mean molecular weight and $k_{B}$ is the Boltzmann constant.
The code numerically solves the equation for the mass conservation:

\begin{equation}\label{eq:rho}
  \pd{\rho}{t} +\nabla \cdot (\rho \mathbf{v}) = 0\,,
\end{equation}

momentum conservation:

\begin{equation}\label{eq:rhov}
   \pd{\rho \vec{v}}{t}
 + \nabla \cdot \bigg[\rho\vec{v}\vec{v} + P_t - \vec{B}\vec{B}\bigg]
 = -\rho\nabla\Phi_{G} = 0 \,,
\end{equation}

magnetic induction (Faraday's law):

\begin{equation}\label{eq:B}
    \pd{\vec{B}}{t}
  - \nabla \times (\vec{v} \times \vec{B}) = 0 \,,
 \end{equation}

and total energy conservation:

\begin{equation} \label{eq:e}
   \pd{e}{t}
 + \nabla \cdot \bigg[\left(e + P_{t}\right)\vec{v}
 - \vec{B}(\vec{v}\cdot \vec{B}) \bigg]
  = - \Lambda_{c} + \mathcal{H}_\X \,.
\end{equation}

In the expressions above, $P_t = p + \vec{B}^2/2$ is the total (gas + magnetic) pressure, $\Phi_{G}$ is the gravitational potential described (see below) while

\begin{equation}
  e = \frac{P}{\gamma-1} + \frac{\rho \vec{v}^2}{2}
                         + \frac{\vec{B}^2}{2} + \rho \Phi_{G} \,,
\end{equation}

is the total energy density with $\Gamma = 5/3$ being the specific heat ratio.
Note that a factor $\sqrt{4\pi}$ has been conveniently re-absorbed in the definition of $\vec{B}$.
We adopt the SNEq (Simplified Non-Equilibrium cooling, see \cite{RO97,TE08}) treatment of the radiative losses. This cooling function includes line emission from nine elements: $\rm{H}$ and $\rm{He}$ resonance lines, and the 13 strongest forbidden lines of $\rm{C, N, O, S, Si, Fe}$ and $\rm{Mg}$, whose abundances have been assumed to be solar.
The temporal evolution of the ionization of $\rm{H}$ is followed by integrating, along with the fluid equations, the evolution equation \ref{eq:fn} for the number fraction of neutral hydrogen atoms $f_n$ \cite{RO97}.
The charge-exchange mechanism with $\rm{H}$ is applied for the other elements. Furthermore we assume that the gas never becomes doubly ionized.
Density sensitivity of the emission lines is considered for forbidden lines having critical densities below $10^5$ cm$^{-3}$.
$\Lambda_{c}$ in Eq. \ref{eq:e} then represents the energy loss term (energy per unit volume per unit time) which includes energy lost in lines and in the ionization and recombination processes.
This cooling treatment is valid for shock velocities below about 80 km s$^{-1}$ and temperatures up to $T \approx 4 \times 10^4$ K.
\cite{TE08} extended the SNEq treatment, where the electrons from hydrogen ionization are considered only, to a more general MINEq (Multi-Ion Non-Equilibrium cooling), where the line emission can be computed in conditions of non-equilibrium ionization for 29 ion species, whose number fraction was obtained by solving the corresponding temporal evolution equations, analogous to Eq. \ref{eq:fn}. Tests performed by the authors have shown that a major advantage obtained by using MINEq, compared to SNEq, was that the line emission could be computed in conditions of non-equilibrium ionization for all species, more likely to be encountered in situations of rapid changes, as it is the case of strong shock waves. However, jet simulations with typical post-shock temperatures in the range $1.5-4 \times 10^4$ K yielded morphologies similar between the SNEq and MINEq runs, since at these temperatures the two cooling loss functions are comparable.

The degree of ionization is governed by an additional evolutionary equation,

\begin{equation}\label{eq:fn}
  \pd{f_n}{t} + (\vec{v} \cdot \nabla) f_{n} = - n_{e}
  \left[\left(c_{i} +\frac{\mathcal{\zeta}_\X}{n_{e}}\right)f_{n}
         + c_{r}(1-f_{n})\right] \,,
\end{equation}

where $f_n=n_{\Hyd \rm I}/n_{\Hyd}$ is the neutral hydrogen fraction, $n_{e}$ is the electron density while $c_{i}$ and $c_{r}$ are respectively, the ionization and recombination rate coefficients \cite{RO97} given by,

\begin{equation}
  \begin{array}{l}
   \DS c_i = \frac{1.08 \times 10^{-8} \sqrt{T}}{(13.6)^{2}} \cdot
                \exp\left(-\frac{157890}{T}\right)
                \, {\rm cm}^{3}\,{\rm s}^{-1} \,,
   \\ \noalign{\medskip}
   \DS c_r = 2.6 \times 10^{-11} \sqrt{T} \, {\rm cm}^{3}\,{\rm s}^{-1} \,,
  \end{array}
\end{equation}

where $T$ is temperature (in Kelvin).
The additional term $\zeta_\X$ accounts for photoionization and it is described below.
The $\mathcal{H}_\X$ term takes into account the energy input by stellar X-ray photons which, in the keV energy range, can potentially result in the production of fast primary photoelectrons after interacting with gas atoms (or molecules).
Primary photoelectrons, in turn, generate a cascade of secondary electrons \cite{GL97}.
We follow the treatment by \cite{SH02} that ignores the contribution of the primary electrons and considers the dominant secondary electrons only.
The X-rays optical depth is

\begin{equation} \label{eq:tauX}
 \tau_\X = \sigma_{pe}(kT_\X) N\,, \qquad N = \int_{0}^{r} n_{H}(r)\, dr\,,
\end{equation}

where $\sigma_{pe}(E) = \sigma_{pe}(kT_\X)(\mbox{keV}/E)^{p}$.
For solar abundances, $p=2.485$, $kT_\X = \mbox{1keV}$ and $\sigma_{pe} (\mbox{1keV}) = 2.27 \times10^{-22} {\rm cm}^{2}$.
We write the energy input $ \mathcal{H}_\X$ by X-rays (energy per unit volume per unit time) and the ionization rate due to the secondary electrons $\zeta_\X$ as

\begin{equation}\label{eq:zetaX}
  \mathcal{\zeta}_\X = \frac{1}{4\pi r^{2}}
  \int_{E_{0}}^{\infty}\frac{L_{\X}(E)}{\epsilon_{\rm ion}}
                       \sigma_{pe}(E) e^{-\tau_\X}\, dE \,.
\end{equation}

In the expression above $L_\X(E)$ is the X-ray luminosity per unit energy interval \cite{SH02}, $E_0(=0.1 \mbox{keV})$ is the low-energy cutoff, $y_{\rm heat}$ is the absorbed fraction of the X-ray flux, and $\epsilon_{\rm ion}$ the energy to make an ion pair.
Since $y_{\rm heat}$ and $\epsilon_{\rm ion}$ can be considered nearly independent of energy, we have

\begin{equation}\label{eq:HX}
  \mathcal{H}_\X = n_\Hyd(r) y_{\rm heat} \epsilon_{\rm ion} \zeta_\X \,
\end{equation}

where \cite{SH85}

\begin{equation}
  \frac{1}{\epsilon_{\rm ion}} = \frac{y_\Hyd}{I_\Hyd} + \frac{y_\Hel}{I_\Hel}\,,
\end{equation}

with
\begin{equation}
  \begin{array}{l}
  y_\Hyd = 0.3908 \left(1 - x_{e}^{0.4092}\right)^{1.7592} \,, \\ \noalign{\medskip}
  y_\Hel = 0.0554 \left(1 - x_{e}^{0.4614}\right)^{1.666} \,.
  \end{array}
\end{equation}

In the above relationships $I_\Hyd$ and $I_\Hel$ are the ionization potentials of $H$ and $He$, $x_{e}$ is the hydrogen ionization fraction and

\begin{equation}
  y_{\rm heat}= 0.9971\left[1 - \left(1 - x_{e}^{0.2663}\right)^{1.3163}\right]
\end{equation}

specifies the heating fraction.
Note that for a thermal spectrum, the ionization rate Eq. (\ref{eq:zetaX}) takes the form

\begin{equation}\label{eq:zetaX_thermal}
  \mathcal{\zeta}_\X = \frac{L_\X\sigma_{pe}(kT_\X)}{4\pi r^2\epsilon_{\rm ion}}
  \int_{\zeta_{0}}^{\infty} \xi^{-p} \exp [-(\xi + \tau_\X\xi^{-p} )] d \xi \,,
\end{equation}

where $L_\X$ is now the total integrated luminosity in $[E_0,\infty]$.
We will also follow the temporal evolution of a passive 'tracer' ${\cal T}$ which is advected by the fluid motion and obeys the simple transport equation

\begin{equation}
 \pd{(\rho {\cal T}_{\rm AJ} )}{t} + \nabla \cdot(\rho {\cal T}_{\rm AJ} \vec{v})=0 \,.
\end{equation}

The addition of a passive scalar will be useful to discriminate between the jet material accelerated due to the magneto-centrifugal mechanism and the jet on the axis, see \S\ref{sec:initial_conditions}.

\subsection{Numerical approach}

The MHD equations~(\ref{eq:rho})--(\ref{eq:e}) together with~(\ref{eq:fn}) are solved in 2D axisymmetric cylindrical coordinates $(R,z)$ using the PLUTO code with adaptive mesh refinement \cite{Mig-PLUTO2012}.
The computational domain is defined by the rectangular box $R\in[0, R_{\rm end}]$, $z\in[0,z_{\rm end}]$, where $R_{\rm end} = 125\,{\rm AU}$ while $z_{\rm end} = 2000 \,{\rm AU}$.
We employ 5 levels of refinement starting from a base grid of $128\times 2048$ zones, yielding an equivalent resolution of $4096\times65536$.
A globally second-order Runge-Kutta scheme with linear spatial reconstruction is used to evolve the equations in time.
Slope limiters in curvilinear coordinates are obtained using the procedure of \cite{Mig2014} to prevent loss of accuracy near the coordinate origin.
The van Leer limiter is applied to primitive flow quantities.
The equations are solved in conservative form by evaluating inter-zone numerical fluxes resulting from the (approximate) solution of Riemann problems at zone interfaces.
For the present work, we employ the Harten-Lax-van Leer (HLL) Riemann solver.
The divergence-free condition of the magnetic field is enforced using hyperbolic divergence cleaning as described in \cite{MigTze2010, Mig-PLUTO2012}.
We employ a combination of static mesh refinement close to the central object, as well as dynamic mesh refinement as the outflow propagates outwards.
More specifically, the region within $20\,{\rm AU}$ is always covered with 4 refinement levels, with one additional level (level 5) in the innermost region ($\lesssim 8\,{\rm AU}$).
Outside of this region but close to the disk boundary ($R > 20\,{\rm AU}$, $z < 5\,{\rm AU}$), the refinement is also static but uses only 3 levels.
Everywhere else the grid is refined dynamically using the second derivative of pressure (although no more than 3 levels are permitted).
We point out that high resolution is essential close to the central object in order to properly resolve the photoionization process, which takes place on a much smaller scale compared to the overall size of the domain.

\subsubsection{Evaluation of the heating Term} \label{Evaluation of the Heating Term}

Evaluation of the heating term, Eq.~(\ref{eq:HX}), requires some additional considerations because the integral in Eq.~(\ref{eq:zetaX_thermal}) depends on the optical depth $\tau_\X$ (Eq.~\ref{eq:tauX}), which in principle should be computed by ray-tracing integration from the central star to the current zone position.
However, in practice the integral in Eq.~(\ref{eq:zetaX_thermal}) becomes negligible at a distance of $\gtrsim 20$\,AU, so that only the statically refined patches around the star yield a significant contribution.
Ray tracing is performed as follows: each zone $(i,j)$ is connected to the origin by a ray divided into $m$ equally spaced segments.
The optical depth is then evaluated using a trapezoidal rule,

\begin{equation}\label{eq:optical_depth}
  N_{ij} \approx \sum_{k=0}^{m-1} \left(\frac{\rho_k + \rho_{k+1}}{2}\right)
                                  \Delta r_{ij} \,,
\end{equation}

where $\rho_k$ is obtained using bi-linear interpolation, $\Delta r_{ij} = r_{ij}/m$ while $r_{ij} = (r_i^2 + z_j^2)^\HALF$ is the spherical radius.
In order to have approximately equal sampling on all rays, we set $m = {\rm int}(r_{ij}/\Delta r_0) + 1$, where $\Delta r_0$ is the diagonal of the first computational cell.
Eq.~(\ref{eq:optical_depth}) can be readily evaluated on computational patches that include the system origin.
Conversely, if a patch does not fulfil this condition, we still use Eq.~(\ref{eq:optical_depth}) to compute the portion of the integral that lies on this grid and then add the contribution coming from the (lower or left) adjacent patch.
Since computations are performed in parallel, the full integral is obtained by joining contributions sequentially over a few time steps as the integration propagates across grid patches.
We have verified (using a single static grid computation) that this does not pose a problem, as the solution reaches a stationary configuration in a few dynamical timescales.
Finally, in order to speed up the computation of the integral in Eq.~(\ref{eq:zetaX_thermal}), we preliminarily obtain a table for different values of $\tau_\X$ and then use a lookup table method as in \cite{Vaidya_etal15} to obtain the value of an arbitrary $\tau_X$.

%%%%%%%%%%%%%%%%%%%%%%%%%%%%%%%%%%%%%%%%%%%%%%%%%%%%%%%%%%%%%%%%%%%%%%%%%
%%%%%%%%%%%%%%%%%%%%%%%%%%%%%%%%%%%%%%%%%%%%%%%%%%%%%%%%%%%%%%%%%%%%%%%%%

\section{Setup}

We now describe the initial and boundary conditions adopted for our numerical simulations.
In what follows we denote with $R$ the cylindrical radius and with $r\equiv\sqrt{R^2+z^2}$ the spherical radius.

\subsection{Initial conditions}
\label{sec:initial_conditions}

In order to set the initial conditions in the computational domain, we first define a gravitational potential by placing a central star with mass $M$ at the coordinate origin

\begin{equation}
  \Phi_{G} = \left\{\begin{array}{ll}
    \DS -\frac{G M}{r_0}\frac{1}{\xi}, & \; \text{if}\quad \xi > 1 \,,
    \\ \noalign{\medskip}
    \DS -\frac{G M}{r_0}\left(
              \frac{7}{4}
              - \frac{7}{8}\xi^2
              + \frac{1}{8}\xi^6 \right) & \; {\rm if} \quad \xi\leq 1 \,.
  \end{array}\right.
\end{equation}

Here $\xi = r/r_0$ is the spherical radius normalized to the reference length $r_0$, which we take to be the inner radius of the Keplerian disk.
The smoothing at $\xi\leq 1$ is introduced to avoid the singularity of the gravitational potential of a point-like mass.
The chosen smoothing sets the gravitational acceleration to zero at the origin while preserving the continuity of the gravitational acceleration and its first derivative at $\xi = 1$.

The density profile is given by

\begin{equation}
  \left(\frac{\rho_{\rm c}}{\rho_{0}} \right)^{\gamma -1} = \left\{\begin{array}{ll}
    \DS \frac{1}{\xi} , & \; {\rm if}\quad \xi > 1
    \\ \noalign{\medskip}
    \DS \left(
              \frac{7}{4}
              - \frac{7}{8}\xi^2
              + \frac{1}{8}\xi^6 \right) & \; {\rm if} \quad \xi \leq 1 \,,
  \end{array}\right.
\end{equation}

where $\rho_0$ is our unit density and $\gamma=\frac{5}{3}$ is the specific heat ratio.
The thermal pressure profile is initially set to $P_{\rm c}= P_0 \left( \rho_{\rm c}/\rho_0\right)^{\gamma}$, so that the initial corona is isentropic. A value $P_0 = (\gamma-1)/\gamma\; GM\rho_0/r_0$ would correspond to a hydrostatic corona. Such a condition would set the initial temperature around the origin close to the virial value ($\sim 10^5 - 10^6$ K for a typical normalization). Since such a high temperature would strongly compromise the computation of the ionization fraction, we instead impose a much lower temperature using $P_0 = 0.0024 \; GM\rho_0/r_0$. The corona is therefore not in hydrostatic equilibrium and tends to collapse towards the origin. This does not pose a problem, as the initial corona is swept up by the formation of the outflow accelerated from the bottom boundary.

While a magneto-centrifugal mechanism is responsible for the disk-wind acceleration, this process cannot be effective around the symmetry axis. To avoid using a thermal pressure gradient (which would introduce unrealistically high temperatures), we added an artificial accelerating force for the axial jet proportional to gravity. We introduced a passive scalar ${\cal T}_{\rm AJ}$ to follow the matter injected from the bottom boundary in the $R<r_0$ region, where ${\cal T}_{\rm AJ}$ is injected with a unity value and zero elsewhere. We then define the gravity as $\vec{g} = (1.2 {\cal T}_{\rm AJ} -1)\nabla \Phi_{G} $, providing an extra push for the axial jet only. In the corona, all velocity components are initially set to zero.

The atmosphere (and the disk) is threaded by a large-scale force-free magnetic field that is purely poloidal \cite{OU97}. The field is prescribed in terms of the $\phi$ component of the vector potential, $\vec{B}= \nabla\times(A_\phi\hvec{e}_\phi)$ where

\begin{equation}
  A_{\phi} = B_{z0} \sqrt{r_0^2 + z_{d}^{2}}
             \frac{\sqrt{R^{2}+(z+z_{d})^{2}}-(z+z_{d})}{R} \,,
\end{equation}

so that $B_z=1/R\partial_R(RA_\phi)$, $B_R= -\partial_z A_{\phi}$.
In the expressions above $z_{d}$ is the disk thickness (set to $r_{0}$), and $B_{z0}$ is the value of the $z$ component of the magnetic field at $(R=r_0, z=0)$.
In order to avoid abrupt changes in the ionization rate, the neutral fraction of hydrogen is set to its equilibrium value.
The temperature of the corona is $\sim 10^{3}$ K at its base and the mean molecular weight is taken to be $\mu = 0.6$.
Finally, to ensure that the density profiles do not fall below observational limits, we impose a floor value $\rho = 10^{-6} \rho_{0}$.

\subsection{Boundary conditions}

The choice of physical boundary conditions is of utmost importance as they determine the final steady-state solution. With the present setup we have to deal with four boundary regions.

\subsubsection{Bottom boundary}

The accretion disk is modelled through the boundary condition at $z \le 0$.
The disk density is set to be proportional to the initial coronal density so that $\rho_{\rm disk} (R) = \chi \; \rho_{c}(R,z=0)$, where $\chi = 100$.
The radial and toroidal magnetic field components $B_R$ and $B_{\phi}$ are set to be continuous across the boundary by linearly extrapolating from the interior zones, while the $B_z$ component is kept at its initial value in order to conserve the magnetic flux through the disk surface.
We require the magnetic field to be frozen into the disk Keplerian rotation, meaning that in a frame locally co-rotating with the disk the electric field $\vec{E}= -\vec{v} \times \vec{B}$ must be zero.
For the toroidal component of the electric field this implies that the poloidal speed must be parallel to the poloidal magnetic field.
We use the fact that the mass-to-magnetic-flux ratio $\eta = \rho v_{p}/B_{p}$ is invariant along poloidal field lines in a stationary state to extrapolate the poloidal speed from the computational domain into the boundary.
The condition on the poloidal component of the electric field provides the boundary condition for the toroidal speed:

\begin{equation}\label{eq:vphi}
  v_{\phi}= v_{K} + \frac{v_{p}}{B_{p}} B_{\phi} \,,
\end{equation}

where the Keplerian velocity profile $v_K$ is recovered by balancing centrifugal and gravitational terms:

\begin{equation}
v_{K} = \left\{\begin{array}{ll}
    \DS \sqrt{\frac{GM}{R}}, & \; \text{if}\quad R > r_0
    \\ \noalign{\medskip}
    \DS \sqrt{\frac{GM}{4r_0}} \sqrt{7\left(\frac{R}{r_0}\right)^2 - 3\left(\frac{R}{r_0}\right)^{6}}
          & \; {\rm if} \quad R\leq r_0 \,.
  \end{array}\right.
\end{equation}

The second term in Eq.~(\ref{eq:vphi}) accounts for the rotationally induced azimuthal component of the magnetic field. When the simulation starts, the rotating disk winds up the poloidal field and induces a toroidal field component. We do not want the toroidal speed to deviate too much from the Keplerian velocity, so we enforce the value of $v_{p}$ such that the toroidal speed is at least half the Keplerian value. The value of $v_{p}$ is also limited to be less than half the local sound speed.

The pressure of the disk is set assuming a thermal height scale in the $R>r_0$ region, defined through the disk aspect ratio $\epsilon = c_{s}/v_{K}$, where $c_{s}^2=p/\rho$ is the disk isothermal sound speed. The disk pressure is therefore given by $P_{\rm disk} = \epsilon^2 \chi \; (\rho_{\rm disk}/\chi\rho_0)^\gamma \; GM\rho_0/r_0$. Note that the disk is also isentropic and its pressure is proportional to the initial coronal pressure at $z=0$. We set the ionization to zero (all neutral).

\subsubsection{Top boundary conditions}

On the outer right boundary ($R=R_{\rm end}$) outflow conditions (zero gradient) are imposed on the poloidal velocity components, neutral fraction of hydrogen, tracer and $B_{r}$.
For density $\rho$, thermal pressure $P$, $v_{\phi}$, $B_{r}$ and $B_{\phi}$ continuity of the first derivative is required.
The condition on $B_{z}$ is determined by imposing the solenoidality condition $\vec{\nabla} \cdot \vec{B} =0$.

\subsubsection{Axial boundary}

Along the jet axis ($R=0$), axisymmetric boundary conditions are applied.
This requires the normal and toroidal components of vector fields to be antisymmetric, while the remaining quantities are symmetric.

\subsubsection{Outer radial boundary}

At the top boundary ($z=z_{\rm end}$), we impose continuity up to the first derivative for $\rho$, $P$, $v_{\phi}$, $B_{z}$ and $B_{\phi}$.
The neutral fraction of hydrogen, tracer, radial and vertical velocities are simply copied from the interior zones (outflow condition).
The normal component of the magnetic field, $B_{r}$, again follows from the divergence-free condition.

\subsubsection{Units}

Fluid variables are normalized to their physical values at the inner disk radius $r_{0}=0.1$ AU, which also sets our unit length.
At that point, the Keplerian velocity is

\begin{equation}
  V_{k0} = 94 \, \bigg(\frac{M}{M \odot} \bigg)^{-1/2}
                 \bigg(\frac{r_{0}}{0.1\; {\rm AU}} \bigg)^{1/2}
   \, {\rm km/s}^{-1} \; .
\end{equation}

The time unit follows from $t_{0}= r_{0}/V_{k0}$,

\begin{equation}
  t_{0} = 1.7 \ \bigg(\frac{M}{M \odot} \bigg)^{-1/2}
                \bigg(\frac{r_0}{\; {\rm AU}} \bigg) ^{3/2}
          \, {\rm days} \; ,
\end{equation}

Densities are given in units of $\rho_0=10^{-18}\,{\rm gr/cm^3}$.

%%%%%%%%%%%%%%%%%%%%%%%%%%%%%%%%%%%%%%%%%%%%%%%%%%%%%%%%%%%%%%%%%%%%%%%%%
\section{Results: adiabatic case}
%
%
%%%%%%%%%%%%%%%%%%%%%%%%%%%%%%%%%%%%%%%%%%%%%%%%%%%%%%%%%%%%%%%%%%%%%%%%%

\begin{figure*}[!h]
  \centering
%\section{Results: Adiabatic Case}%
 \includegraphics[width=0.33\textwidth]{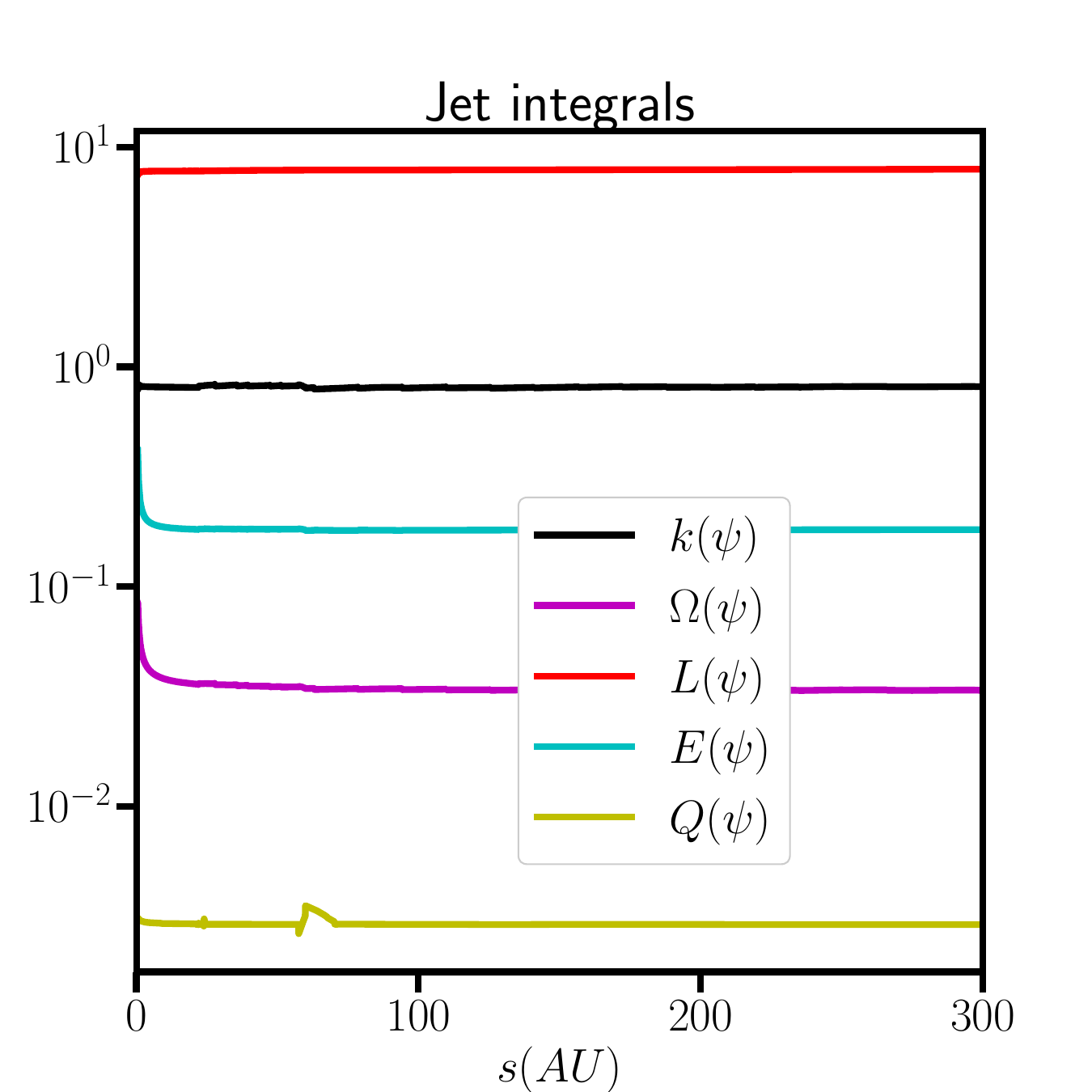}
  \includegraphics[width=0.33\textwidth]{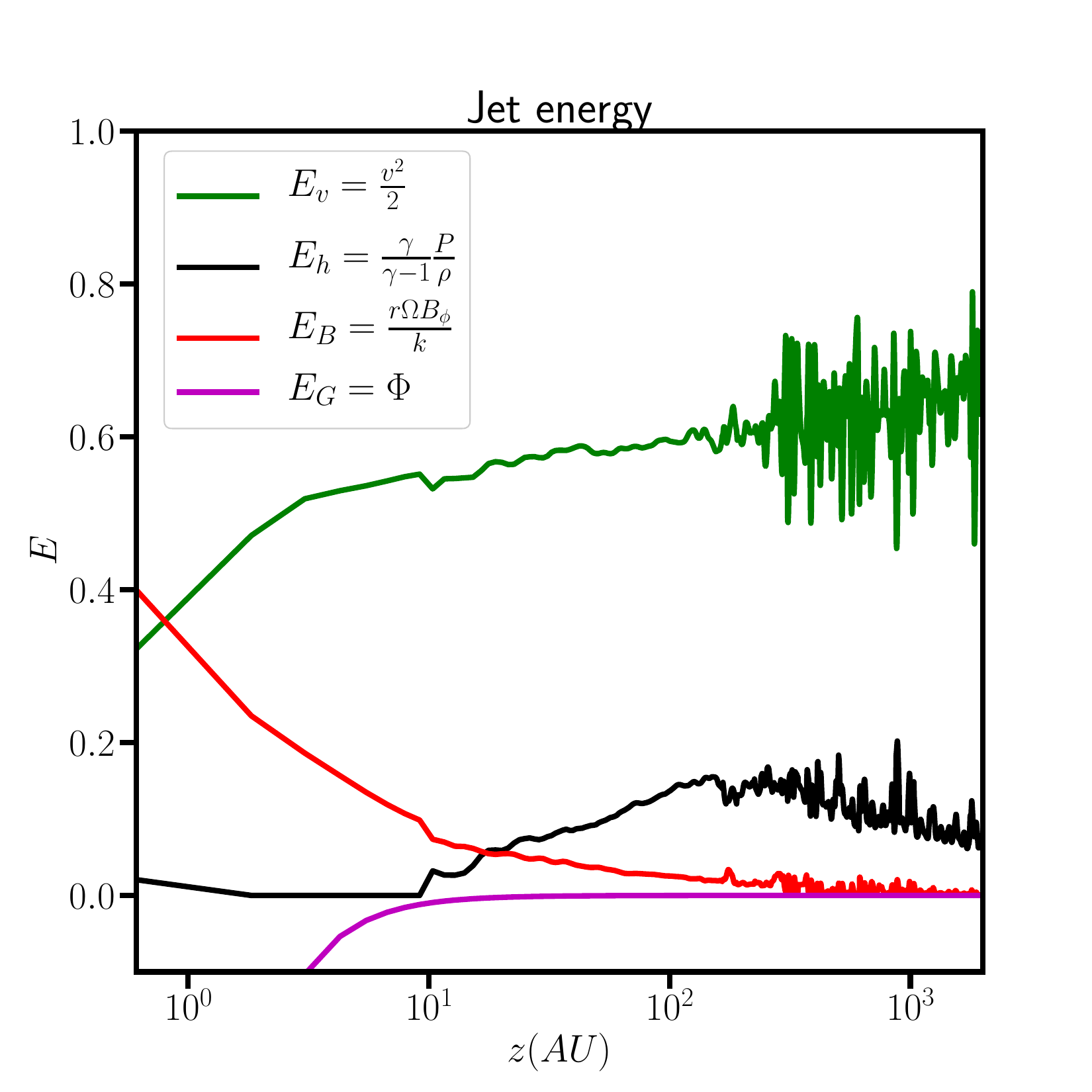}%
  \includegraphics[width=0.33\textwidth]{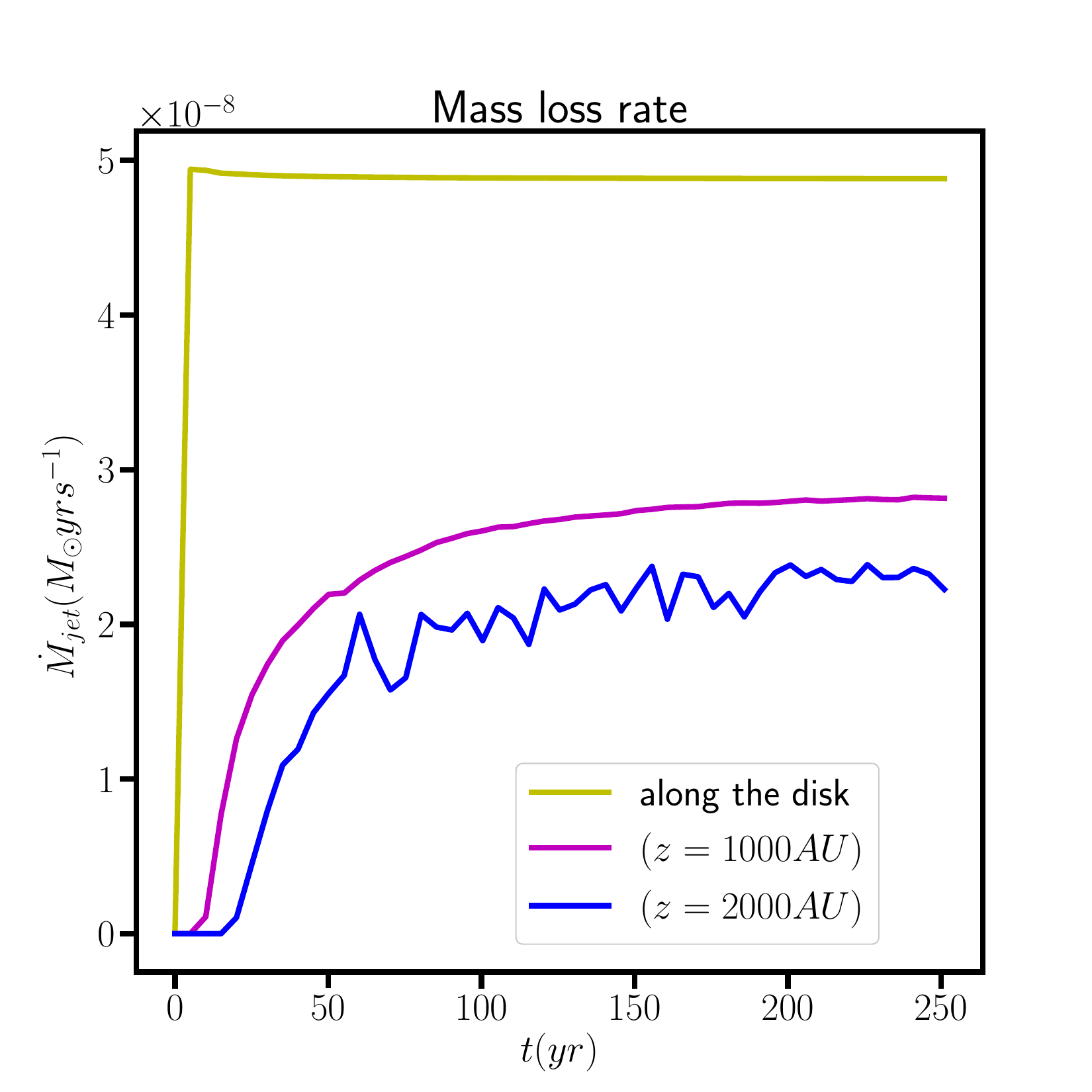} 

  \caption {\footnotesize Left panel: Jet integrals along the magnetic 
  field line rooted in the innermost disk. 
  The jet specific angular momentum $\Omega$, the mass load $k$, the 
  field angular velocity $L$, the jet specific energy $E$, and the specific 
  entropy $Q$. 
  Central panel: Jet specific energy contributions along the vertical 
  direction $z$. 
  Different energy components are indicated by colors:  kinetic (green), 
  magnetic (red), gravitational (magenta), and thermal (black) energy. 
  Right panel: Mass loss rate as a function of time  at different heights (yellow) along the disk, (magenta) at $1000$AU and (blue) at $2000$AU. The figure show decrease in the ejection as we go further from the disk surface.
  } 
\label{fig:streamline_integral}
\end{figure*}

In this section we check our numerical results against theoretical predictions by using the adiabatic MHD simulation (no cooling or heating).

%%%%%%%%%%%%%%%%%%%%%%%%%%%%%%%%%%%%%%%%%%%%%%%%%%%%%%%%%%%%%%%%%%%%%%%%%
\subsection{Steady state jet MHD integrals}
%%%%%%%%%%%%%%%%%%%%%%%%%%%%%%%%%%%%%%%%%%%%%%%%%%%%%%%%%%%%%%%%%%%%%%%%%
In axisymmetric, stationary, ideal MHD there are five quantities that are expected to be conserved along a field line, i.e., a surface of constant magnetic flux $\Psi$, namely the mass to magnetic flux ratio, %mass loading per flux surface enclosing the magnetic flux  $\Psi$, \BLACK
%any given magnetic flux line $s$  $ \Psi \equiv \int \mathbf{B}_{p} \cdot  d\mathbf{A}$, 
%
\begin{equation}
 k(\Psi) =\frac{\rho v_{p}}{B_{p}},
\end{equation}
the angular velocity of the field lines:
\begin{equation}
  \Omega( \Psi)= \frac{1}{r} \left(v_{\phi} -\frac{k B_{\phi}}{\rho}\right)\,
\end{equation}
the specific angular momentum
\begin{equation}   
  L(\Psi)=L_{v}+L_{B}= r v_{\phi}-r \frac{B_{\phi}}{k}  \,,   
\end{equation}
with kinematic and magnetic contribution, and the specific energy
\begin{equation}
  E(\Psi)=    \frac{v^{2}}{2} 
            + \frac{\gamma}{\gamma-1} \frac{P}{\rho}
            + \Phi -\frac{r\Omega B_{\phi}}{k} \,,
            \label{specific}
\end{equation}
and the specific entropy,
\begin{equation}
  Q(\Psi)= \frac{P}{\rho^{\gamma}}
\end{equation}

The left panel in Fig. \ref{fig:streamline_integral} displays the values of these integrals as a function of position $s$ along the field line at some advanced stage, when a stationary configuration has been reached.
We use a sample field line rooted in the innermost part of the ejection region at a radius $1.5 r_{0}$ and plot the quantities at $t = 250\, {\rm yrs}$. 
The integrals are approximately constant, as expected, showing small deviation with errors that remain below $6 \%$.
 
%%%%%%%%%%%%%%%%%%%%%%%%%%%%%%%%%%%%%%%%%%%%%%%%%%%%%%%%%%%%%%%%%%%%%%%%%
\subsection{Jet energy}
%%%%%%%%%%%%%%%%%%%%%%%%%%%%%%%%%%%%%%%%%%%%%%%%%%%%%%%%%%%%%%%%%%%%%%%%%
In the central panel of Fig. \ref{fig:streamline_integral} we plot the different contributions to the specific energy of Eq. (\ref{specific}) along the jet vertical direction.
The thermal energy is negligible everywhere in the jet while gravitational energy is negligible only far from the disk.
The plots further indicates how the transition from a magnetically to kinetically dominated flow takes place at a distance $\lesssim 10$ AU.
Far from the disk the transformation of the magnetic energy into kinetic energy becomes less efficient.
This decrease of acceleration is related to the increase of collimation.

\begin{figure*}[!ht]
  \includegraphics[trim=0 0 0 0,clip,width=\textwidth]{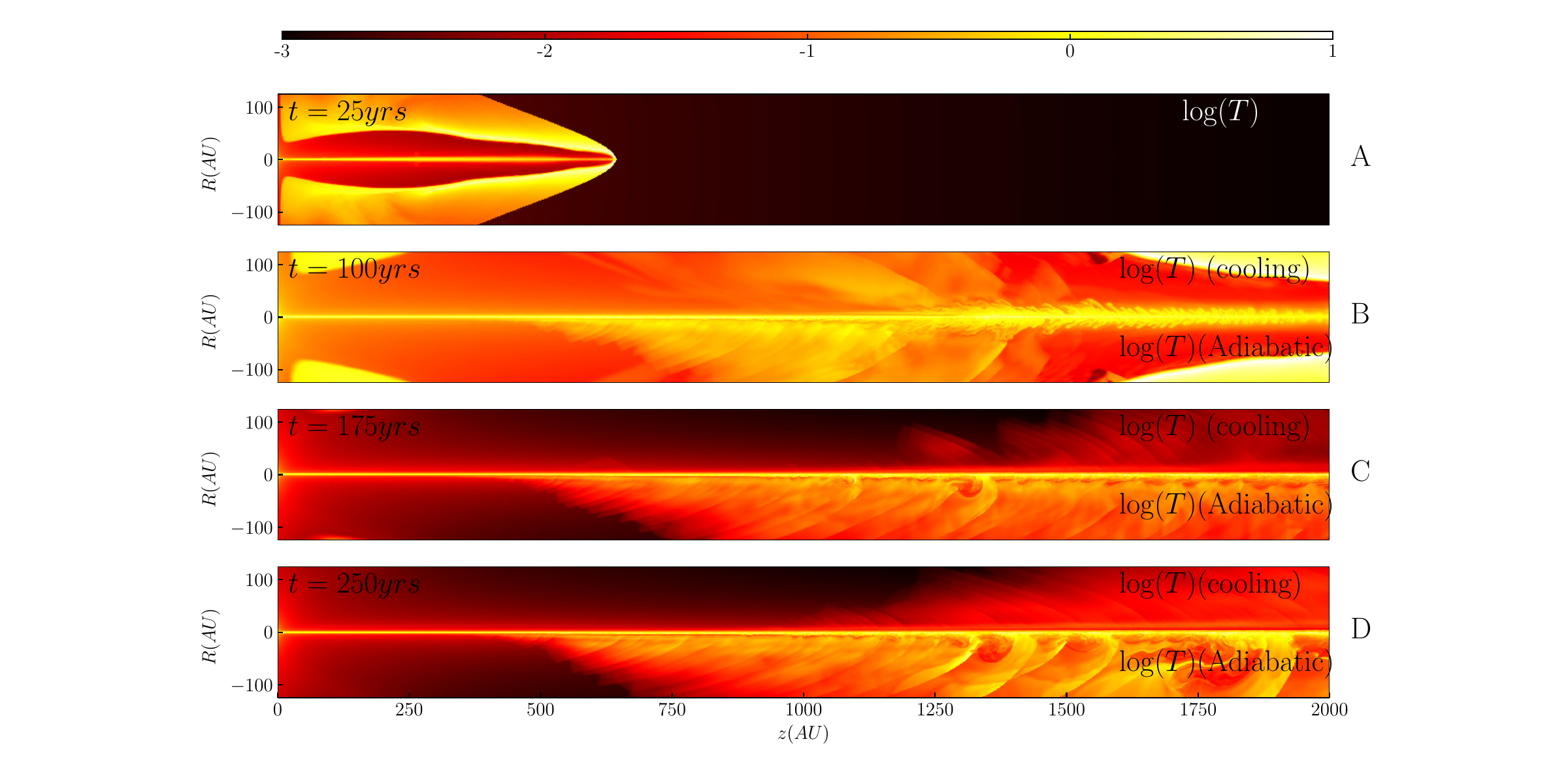}%mapfr
  \caption{\footnotesize From top to bottom, we show four snapshots of the temporal evolution corresponding to $t = 25,100,175 $ and $250\, {\rm yrs}$. 
  In the first panel (A) shows the logarithmic maps 
  of the temperature ( in units of $10^{4}$ K) 
   for the adiabatic case.
  The second three panels (B, C, D) compares the temperature maps when cooling is
  included (upper half) and the adiabatic case (lower half).
  The jet propagates from left to right.}
 \label{fig:Temp_vs_density}
\end{figure*}
%\BLACK
%%%%%%%%%%%%%%%%%%%%%%%%%%%%%%%%%%%%%%%%%%%%%%%%%%%%%%%%%%%%%%%%%%%%%%%%%
\subsection{Mass loss rate}
%%%%%%%%%%%%%%%%%%%%%%%%%%%%%%%%%%%%%%%%%%%%%%%%%%%%%%%%%%%%%%%%%%%%%%%%%

The mass loss rate is a parameter which is, in principle, accessible by observation.
Thus the normalisation of density $\rho_{0} = 10^{-18}\,{\rm gr/cm^3}$ is chosen by setting suitable loss rates   $\rm{\dot{M}_{0}} = r_0^{2} \rho_{0}V_{k0}$: 

\begin{eqnarray}  
 \rm{\dot{M}_{0} = {\rm 4.3 \times 10^{-8}} \ \left(\frac{\rm\rho_{0}}{  10^{-18} {\mathrm {g \ cm^{-3} } }} \right)   \left(\frac{M}{M_\odot} \right)^{1/2} \times  \nonumber } \\  \left( \frac{r_0}{\rm AU} \right)^{3/2} 
 {\mathrm {M_\odot yrs}^{-1}} \;  ,
\end{eqnarray}
%\begin{eqnarray}  
 %\rm{\dot{M_{0}} = {\rm 4.3 \times 10^{-8}} \ \left(\frac{\rm\rho_{0}}{  10^{-18} {\mathrm {g \ cm^{-3} } }} \right)   \left(\frac{M}{M_\odot} \right)^{1/2} \times \nonumber \\
%\left( \frac{r_0}{\rm AU} \right)^{3/2} 
%{\mathrm {M_\odot yrs}^{-1}} \; ,
%\end{eqnarray}
%
We calculate the mass loss rate at different heights in the jet as function of time. 
To this aim, we consider the mass flowing through a surface $S_d = \pi R^2_{\rm end}$ centered around the symmetry axis:
\begin{equation}    
  \dot{M}_{\rm jet} = \int_{S_{d}} \rho \mathbf{v} \cdot d \mathbf{S}     
\end{equation}
The results obtained are shown in the rightmost panel of Fig \ref{fig:streamline_integral} indicating that the mass loss rate is proportional to $10^{-8} M_{\odot} \, {\rm yrs^{-1}}$ and decreases as we move far away from the accretion disk due to losses sideways. The asymptotic values in time at different heights are consistent with observations. 

%%%%%%%%%%%%%%%%%%%%%%%%%%%%%%%%%%%%%%%%%%%%%%%%%%%%%%%%%%%%%%%%%%%%%%%%%
%%%%%%%%%%%%%%%%%%%%%%%%%%%%%%%%%%%%%%%%%%%%%%%%%%%%%%%%%%%%%%%%%%%%%%%%%

\section{Results: adiabatic Case}

\begin{figure*}[!h]
  \centering
 \includegraphics[width=0.33\textwidth]{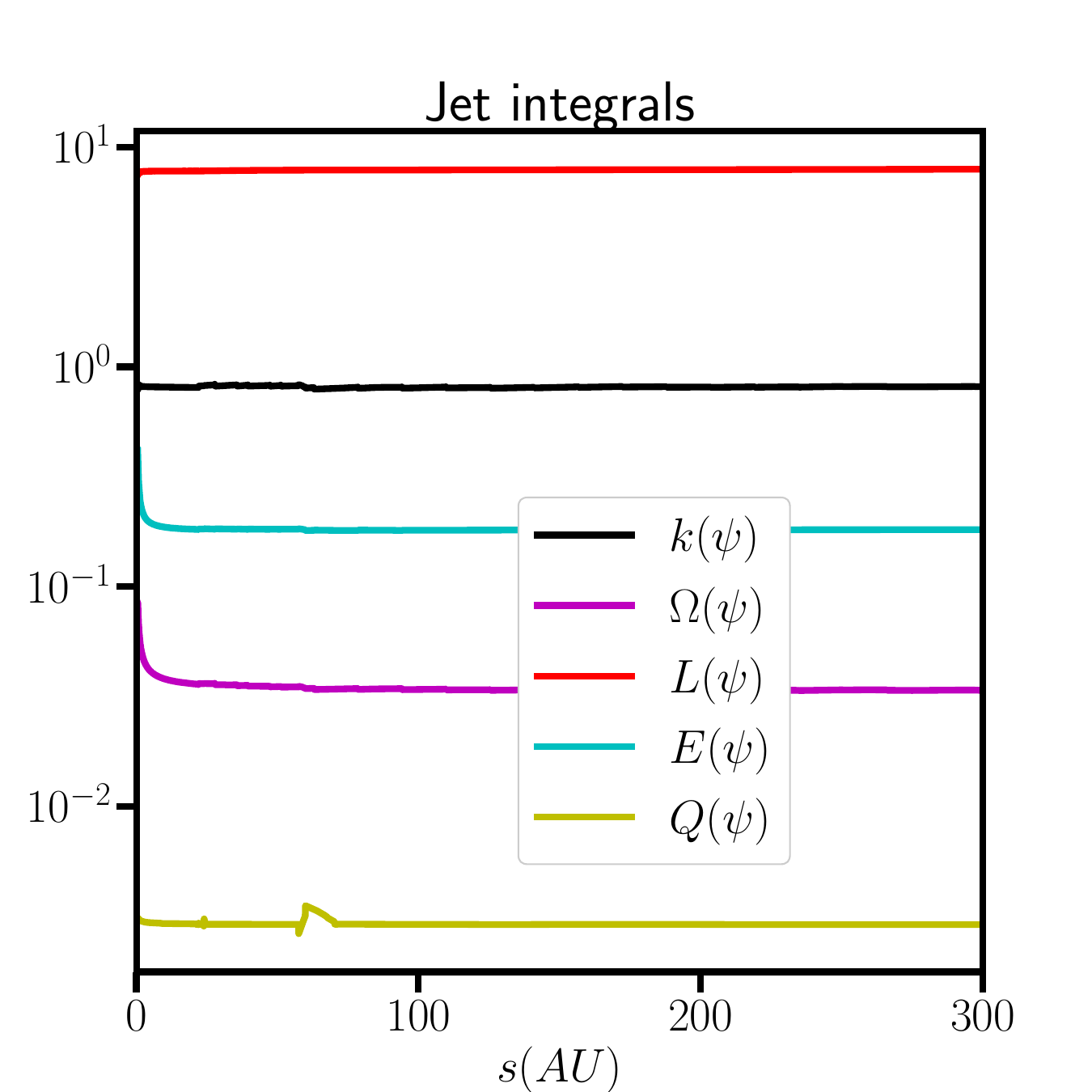}
  \includegraphics[width=0.33\textwidth]{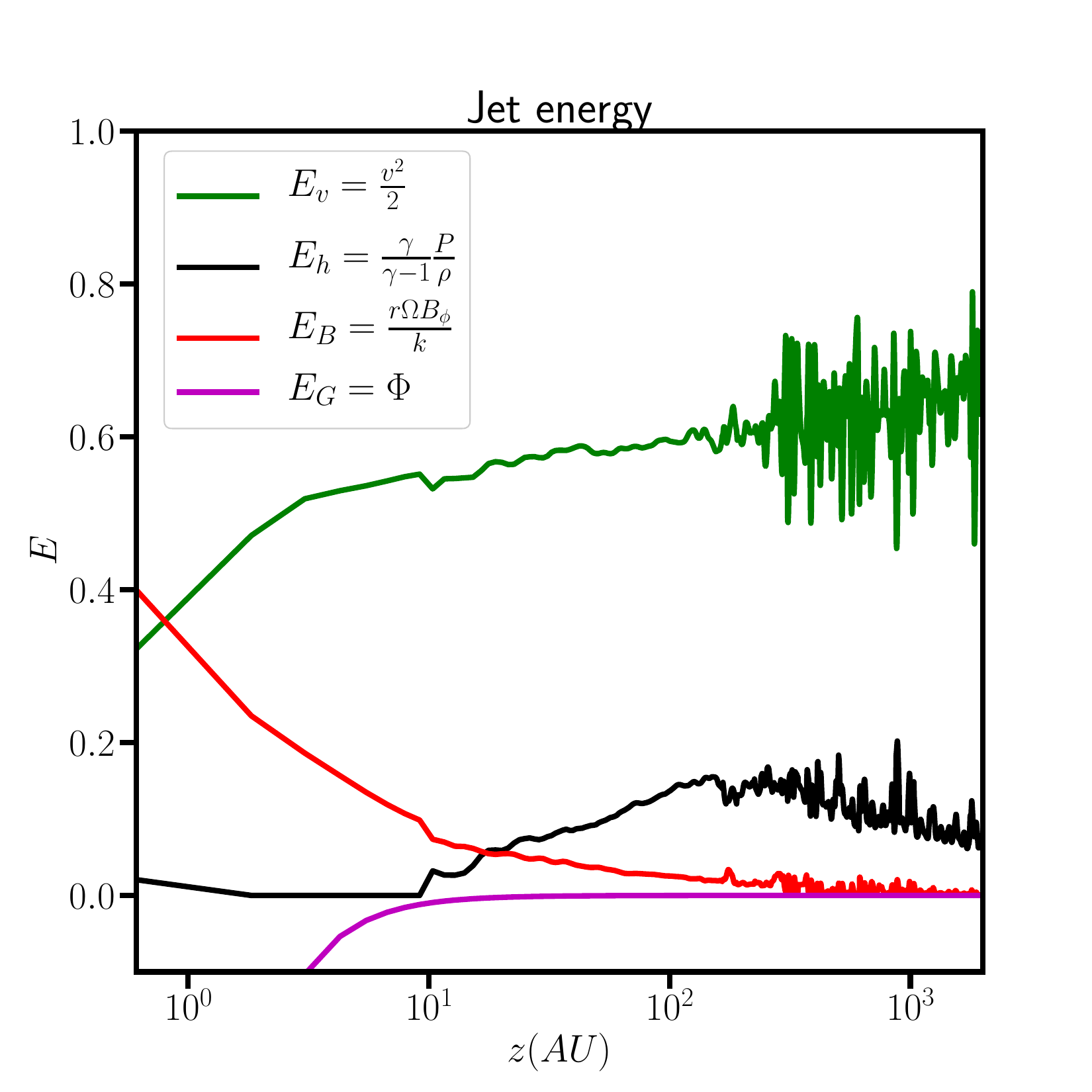}%
  \includegraphics[width=0.33\textwidth]{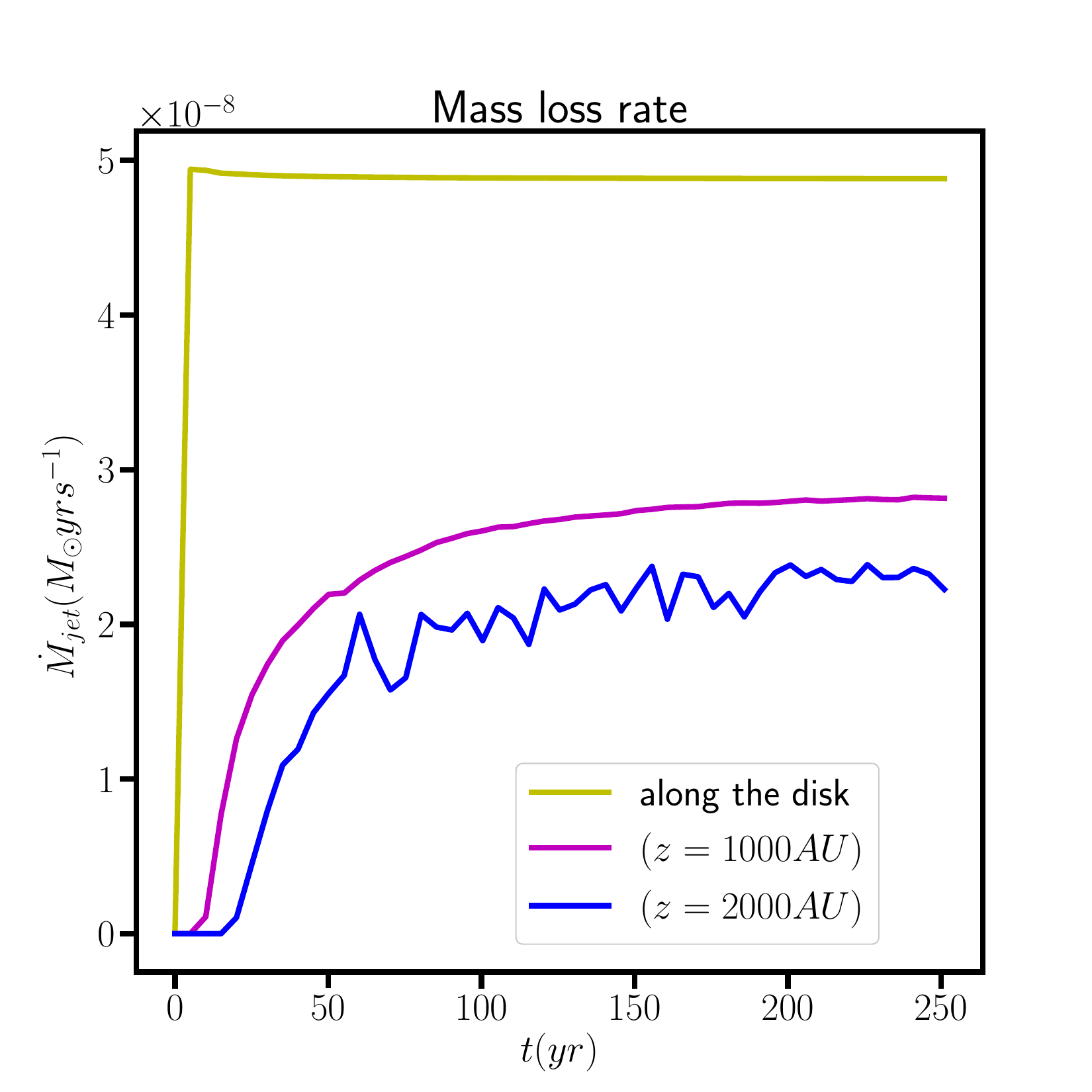}
  \caption {\footnotesize Left panel: Jet integrals along the magnetic
  field line rooted in the innermost disk.
  The jet specific angular momentum $\Omega$, the mass load $k$, the
  field angular velocity $L$, the jet specific energy $E$, and the specific
  entropy $Q$.
  Central panel: Jet specific energy contributions along the vertical
  direction $z$.
  Different energy components are indicated by colors: kinetic (green),
  magnetic (red), gravitational (magenta), and thermal (black) energy.
  Right panel: Mass loss rate as a function of time at different heights (yellow) along the disk, (magenta) at $1000$AU and (blue) at $2000$AU. The figure show decrease in the ejection as we go further from the disk surface.
  }
\label{fig:streamline_integral}
\end{figure*}

In this section we compare our numerical results with theoretical predictions using an adiabatic MHD simulation (no cooling or heating).

\subsection{Steady state jet MHD integrals}

In axisymmetric, stationary, ideal MHD there are five quantities that are expected to be conserved along a field line, i.e., a surface of constant magnetic flux $\Psi$. These are the mass-to-magnetic-flux ratio

\begin{equation}
 k(\Psi) =\frac{\rho v_{p}}{B_{p}},
\end{equation}

the angular velocity of the field lines:

\begin{equation}
  \Omega( \Psi)= \frac{1}{r} \left(v_{\phi} -\frac{k B_{\phi}}{\rho}\right)\,,
\end{equation}

the specific angular momentum

\begin{equation}
  L(\Psi)=L_{v}+L_{B}= r v_{\phi}-r \frac{B_{\phi}}{k} \,,
\end{equation}

with kinematic and magnetic contributions, and the specific energy

\begin{equation}
  E(\Psi)= \frac{v^{2}}{2}
            + \frac{\gamma}{\gamma-1} \frac{P}{\rho}
            + \Phi -\frac{r\Omega B_{\phi}}{k} \,,
            \label{specific}
\end{equation}

and the specific entropy

\begin{equation}
  Q(\Psi)= \frac{P}{\rho^{\gamma}}.
\end{equation}

The left panel in Fig.~\ref{fig:streamline_integral} displays the values of these integrals as a function of position $s$ along the field line at an advanced stage when a stationary configuration has been reached. We use a sample field line rooted in the innermost part of the ejection region at a radius $1.5 r_{0}$ and plot the quantities at $t = 250\, {\rm yrs}$. The integrals are approximately constant, as expected, showing small deviations with errors that remain below $6\,\%$.

\subsection{Jet energy}

In the central panel of Fig.~\ref{fig:streamline_integral} we plot the different contributions to the specific energy of Eq.~(\ref{specific}) along the vertical direction of the jet. The thermal energy is negligible everywhere in the jet, while gravitational energy is negligible only far from the disk. The plots further indicate that the transition from a magnetically to a kinetically dominated flow takes place at a distance $\lesssim 10$ AU. Far from the disk the transformation of magnetic energy into kinetic energy becomes less efficient. This decrease in acceleration is related to the increase in collimation.

\begin{figure*}[!ht]
  \includegraphics[trim=0 0 0 0,clip,width=\textwidth]{denttemp.eps}%mapfr
  \caption{\footnotesize From top to bottom, we show four snapshots of the temporal evolution corresponding to $t = 25,100,175 $ and $250\, {\rm yrs}$.
  In the first panel (A) shows the logarithmic maps
  of the temperature ( in units of $10^{4}$ K)
   for the adiabatic case.
  The second three panels (B, C, D) compares the temperature maps when cooling is
  included (upper half) and the adiabatic case (lower half).
  The jet propagates from left to right.}
 \label{fig:Temp_vs_density}
\end{figure*}

\subsection{Mass loss rate}

The mass loss rate is a parameter which is, in principle, accessible by observation. Thus the normalisation of density $\rho_{0} = 10^{-18}\,{\rm gr/cm^3}$ is chosen by setting suitable loss rates $\rm{\dot{M}_{0}} = r_0^{2} \rho_{0}V_{k0}$:

\begin{eqnarray}
 \rm{\dot{M}_{0} = {\rm 4.3 \times 10^{-8}} \ \left(\frac{\rm\rho_{0}}{ 10^{-18} {\mathrm {g \ cm^{-3} } }} \right) \left(\frac{M}{M_\odot} \right)^{1/2} \times \nonumber } \\ \left( \frac{r_0}{\rm AU} \right)^{3/2}
 {\mathrm {M_\odot yrs}^{-1}} \; ,
\end{eqnarray}

We calculate the mass loss rate at different heights in the jet as a function of time. To this aim, we consider the mass flowing through a surface $S_d = \pi R^2_{\rm end}$ centered around the symmetry axis:

\begin{equation}
  \dot{M}_{\rm jet} = \int_{S_{d}} \rho \mathbf{v} \cdot d \mathbf{S}
\end{equation}

The results obtained are shown in the rightmost panel of Fig.~\ref{fig:streamline_integral}, indicating that the mass loss rate is proportional to $10^{-8} M_{\odot} \, {\rm yrs^{-1}}$ and decreases as we move far away from the accretion disk due to sideways losses. The asymptotic values in time at different heights are consistent with observations.

\section{Results: non-adiabatic cases}

In the non-adiabatic case, cooling and heating are both included in the energy equation~(\ref{eq:e}) and the degree of ionization is computed by solving Eq.~(\ref{eq:fn}). Still, in order to avoid excessive cooling at the jet bow shock (which would cause a strong decrease of the numerical time step), we suppress cooling for $t \lesssim 40\,{\rm yrs}$ of evolution and then gradually turn it on.

\subsection{Dynamics}

The temporal evolution is described by the four panels in Fig.~\ref{fig:Temp_vs_density}, showing temperature maps at four different stages of the evolution. For comparison, the lower panels (B, C, D) show the temperature distributions in the adiabatic and non-adiabatic cases. During propagation the jet bow shock is heated to large temperatures ($T\sim 7 \times10^{4}$ K) and unstable flow patterns develop close to the axis. These perturbations are induced by the combined action of Kelvin-Helmholtz and pressure-driven $m=0$ modes and tend to grow sideways, steepening into oblique shock waves around $t\approx 100\, {\rm yrs}$. While in the adiabatic case these features persist until later times, they become progressively weaker in the presence of radiative losses (see the third and fourth panels in the figure). A steady state is finally reached around $t\approx 250 \,{\rm yrs}$. The effect of cooling is also to reduce the internal energy of the flow by approximately one order of magnitude when compared to the adiabatic run. Owing to the drop in thermal pressure, the jet is more easily collimated towards the axis by the hoop stresses.

\begin{figure}[!h]
  \centering
  \includegraphics[width=0.48\textwidth]{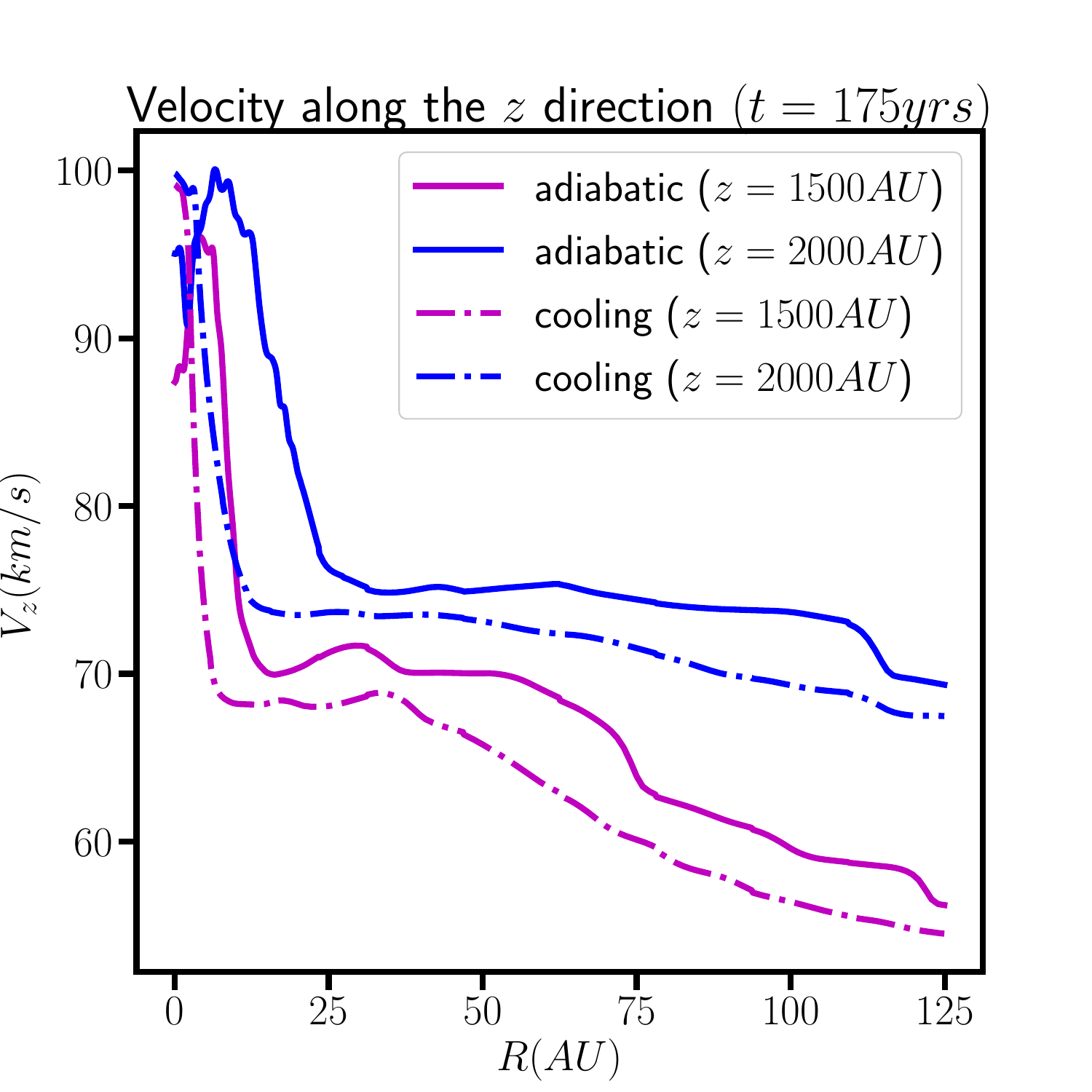}
   \includegraphics[width=0.48\textwidth]{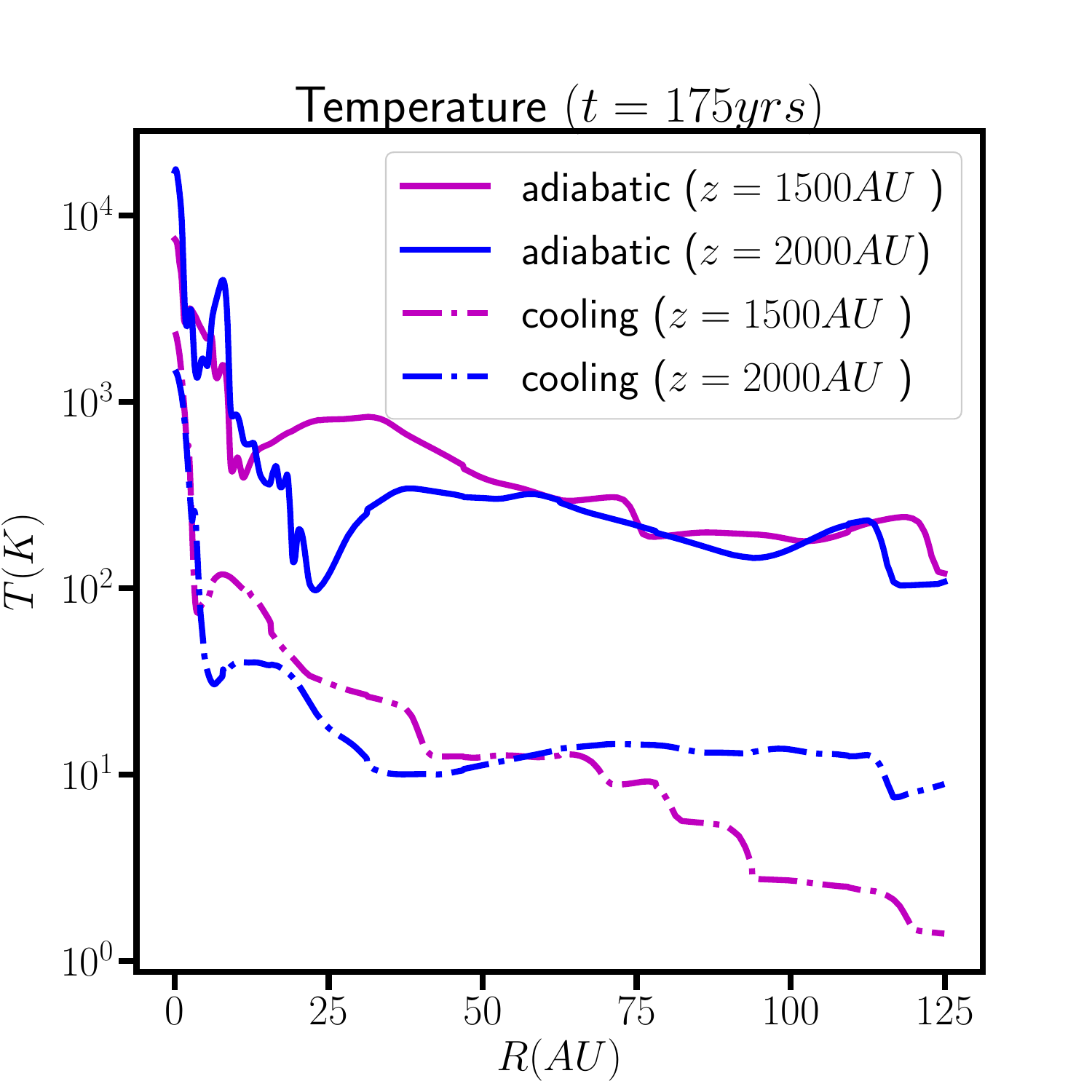}
  \caption {\footnotesize Top panel: velocity along
 the $z$ direction in km/s profile - adiabatic and cooling cases, - at different heights.
  Bottom panel: temperature profile (in $K$) - adiabatic and cooling cases, - at different heights, corresponding to $t = 175 \, {\rm yrs}$. }
  \label{fig:cooling_vs_adiab}
\end{figure}

The effect of energy losses on the jet dynamics is shown in Fig.~\ref{fig:cooling_vs_adiab} where we compare radial profiles of temperature and vertical velocity at different heights of the domain for both the adiabatic and non-adiabatic simulations. Optically thin radiative losses significantly reduce the jet temperature, by one order of magnitude for $R \lesssim 25$ AU and even more outside of this region. The loss of internal energy reduces the efficiency of the jet thrust and the bulk kinetic energy diminishes. As a consequence, the beam velocity is also reduced (lower panel). The velocity indicates that there is a narrow central spine at high velocity surrounded by an extended low velocity wind. The observations often show the same structure (see \cite{ss}).

\begin{figure*}[!ht]
  \includegraphics[width=\textwidth]{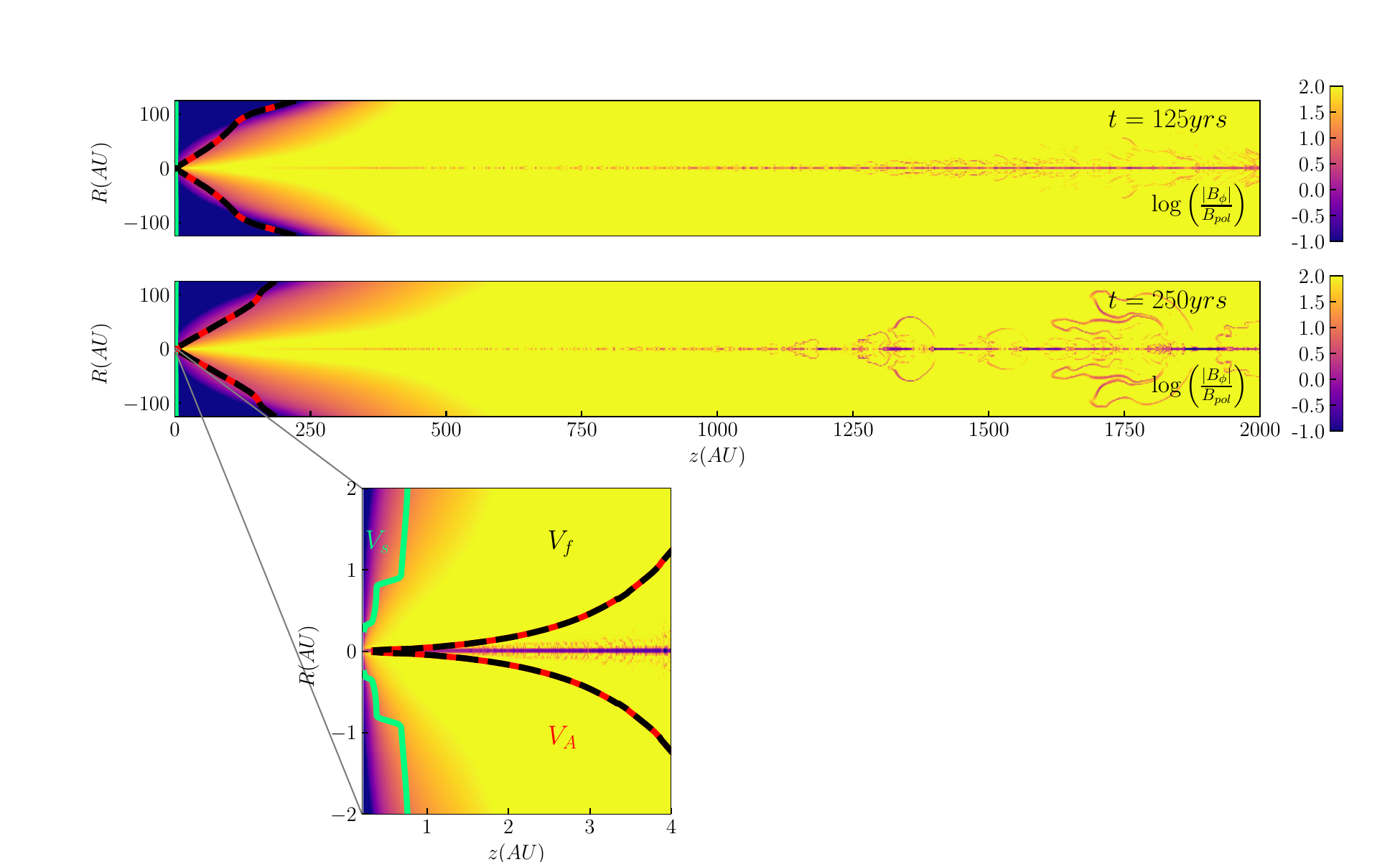}%mapfr
  \caption{\footnotesize Top panel: From top to bottom, we show two snapshots of the logarithmic ratio of the absolute value of the toroidal magnetic field over the poloidal, the temporal evolution corresponding to $t = 125$ and $250\, {\rm yrs}$. The solid lines in (red, black and green) indicate the critical surfaces i.e. Alfv\'{e}nic, fast magnetosonic and slow magnetosonic respectively. Bottom panel: Zoomed image of the jet base, with logarithmic coordinate along $z$ indicating the critical surfaces; the solid lines in (red, black and green) corresponds to Alfv\'{e}nic, fast magnetosonic and slow magnetosonic respectively. The jet propagates from left to right. }
 \label{fig:bpol}
\end{figure*}

\subsubsection{Magnetic field configuration}

The magnetic field configuration, showing the logarithmic ratio of toroidal to poloidal magnetic field components as well as the critical surfaces, is shown in Fig.~\ref{fig:bpol}. Critical surfaces are defined as the isosurfaces where the poloidal velocity component becomes equal to one of the three MHD wave speeds:

\begin{equation}
  \begin{array}{l}
    \DS V_{A}^{2} = \frac{B^{2}}{\rho} \\ \noalign{\medskip}
    \DS V_{f,s}^{2} = \frac{ (C_{s}^{2}+V_{A}^{2} )
                      \pm\sqrt{ (C_{s}^{2}+V_{A}^{2} )^{2}
                               - 4 C_{s}^{2} V_{Ap}^{2} } }
                          { 2}\,,
  \end{array}
\end{equation}

where $V_{Ap}$ is the (poloidal) Alfvénic speed and $C_{s} = \sqrt{\gamma p/\rho}$ is the adiabatic speed of sound. The three isosurfaces are plotted as solid lines (red, black and green) corresponding to Alfvénic, fast magnetosonic and slow magnetosonic, respectively. The flow crosses the Alfvénic surface when the bow shock has left the domain and the system has entered a stationary configuration. The colormap clearly shows that there are two regions where the poloidal component exceeds the toroidal one, namely near the disk surface and below the Alfvén surface (depicted in cyan solid line). Close to the disk the poloidal component is dominant to impose the co-rotation of the outflow. Conversely, above the Alfvén surface co-rotation stops and the azimuthal component dominates, so that the main driving force is the gradient of magnetic pressure of $B_{\phi}$. For a complete description of the mechanism see \cite{BL82}.

\subsection{Ionization}

Next we investigate the effect of cooling and heating on the amount of ionization produced during the evolution. To this purpose, we compare two different simulation cases: the former including the optically thin losses alone and the latter also introducing the heating term due to photoionization.

\begin{figure*}[!ht]
  \includegraphics[trim=0 60 40 0,clip,width=0.9\textwidth]{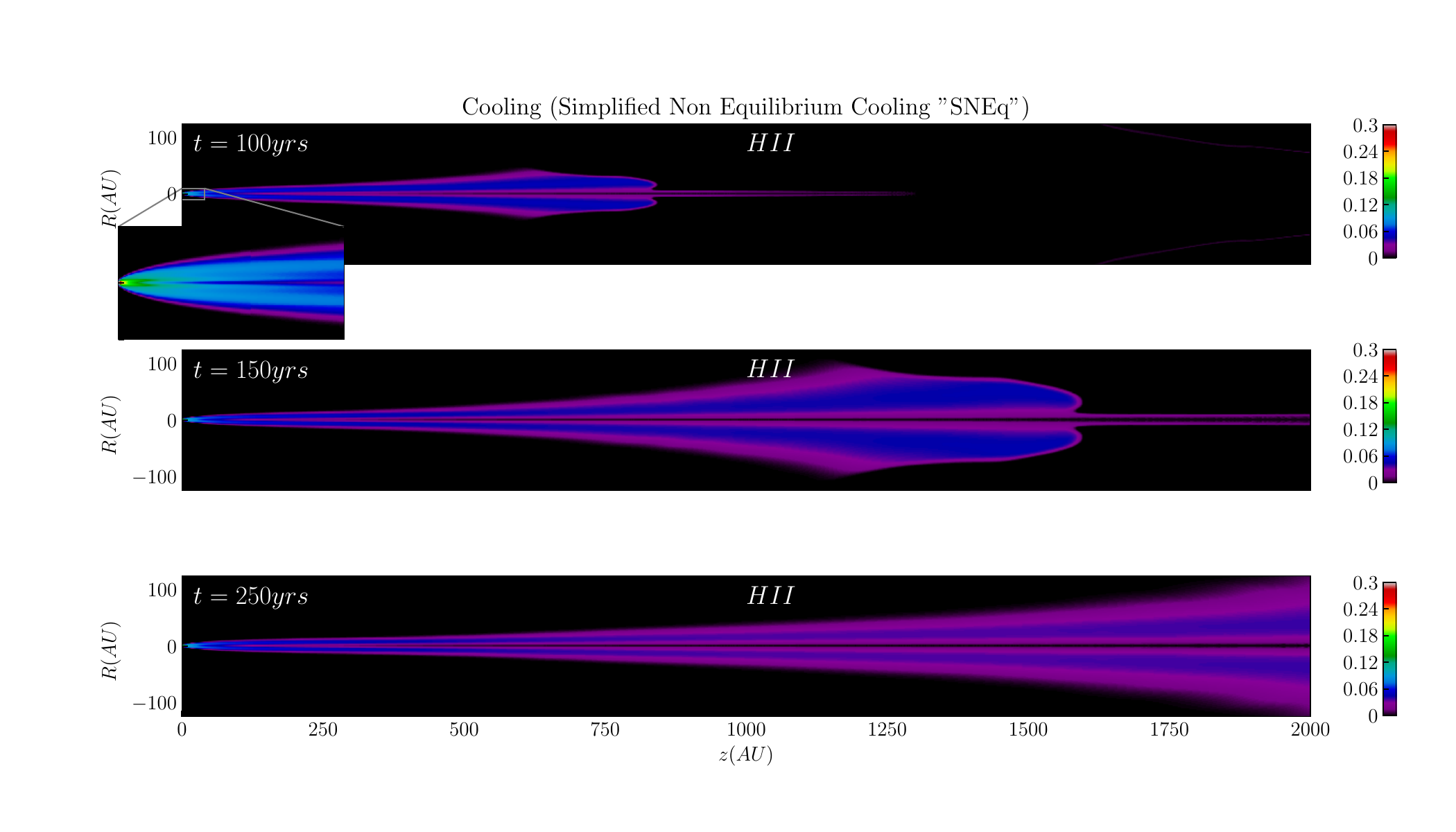}
  \caption{\footnotesize SNEq (Simplified Non Equilibrium Cooling): Ionization fraction of the jet material when
   only cooling is included (no heating).
   Snapshots are shown at $t= 100, 150$ and $t= 250\, {\rm yrs}$.
}
  \label{fig:ionization_fraction_SNEq}
\end{figure*}

In the first case, only $\sim 10\,\%$ of the hydrogen becomes ionized close to the central object and recombination takes place as the flow propagates outward, eventually lowering this value to less than $4\,\%$ at large distances. This situation is best illustrated in Fig.~\ref{fig:ionization_fraction_SNEq} where we display 2D maps of the ionization fraction at $t=100$, $150$ and $250\,{\rm yrs}$, when steady state is eventually reached. The ionization area is initially small ($\sim 2$ AU) and it expands conically, reaching $\sim 120$ AU at the end of the domain. These results are consistent with the idea that the ionisation persists while travelling away from the source \cite{BE99}.

\begin{figure*}[!ht]
  \includegraphics[width=0.9\textwidth]{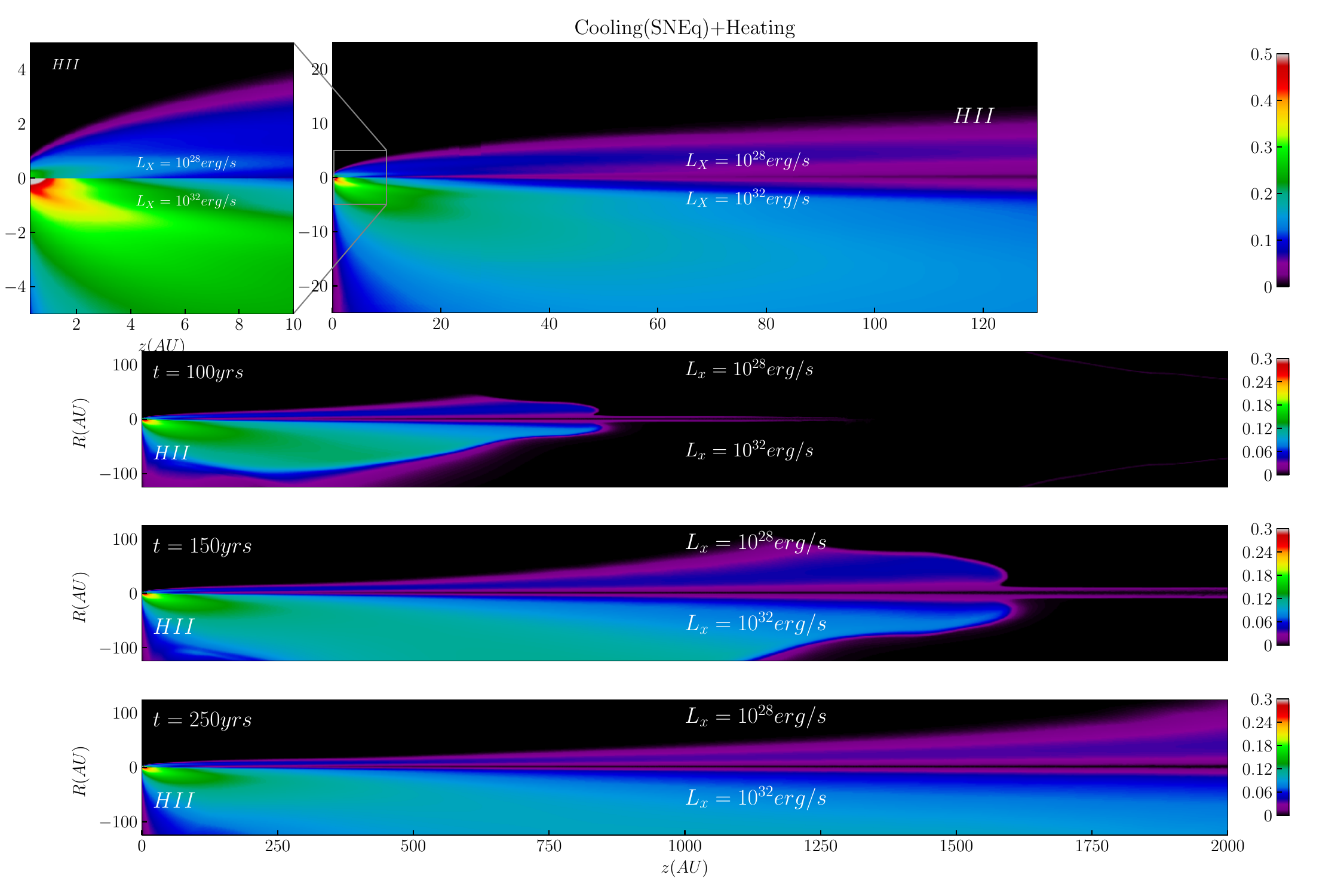}%mapfr
  \caption{\footnotesize
   (SNEq+Heating) Bottom panel: Maps of the total ionization fraction of
   the jet material for different luminosity $L_{\X} = 10^{28}$ erg/s,
   $L_{\X} = 10^{32}$ erg/s at different times $t=100, 150$ and $t= 250\, {\rm yrs}$. Top panel: The corresponding zoomed image of total hydrogen ionization fraction at time $t=100\, {\rm yrs}$. }
  \label{fig:ionization_fraction}
\end{figure*}

\begin{figure}[!h]
  \centering
  \includegraphics[width=0.48\textwidth]{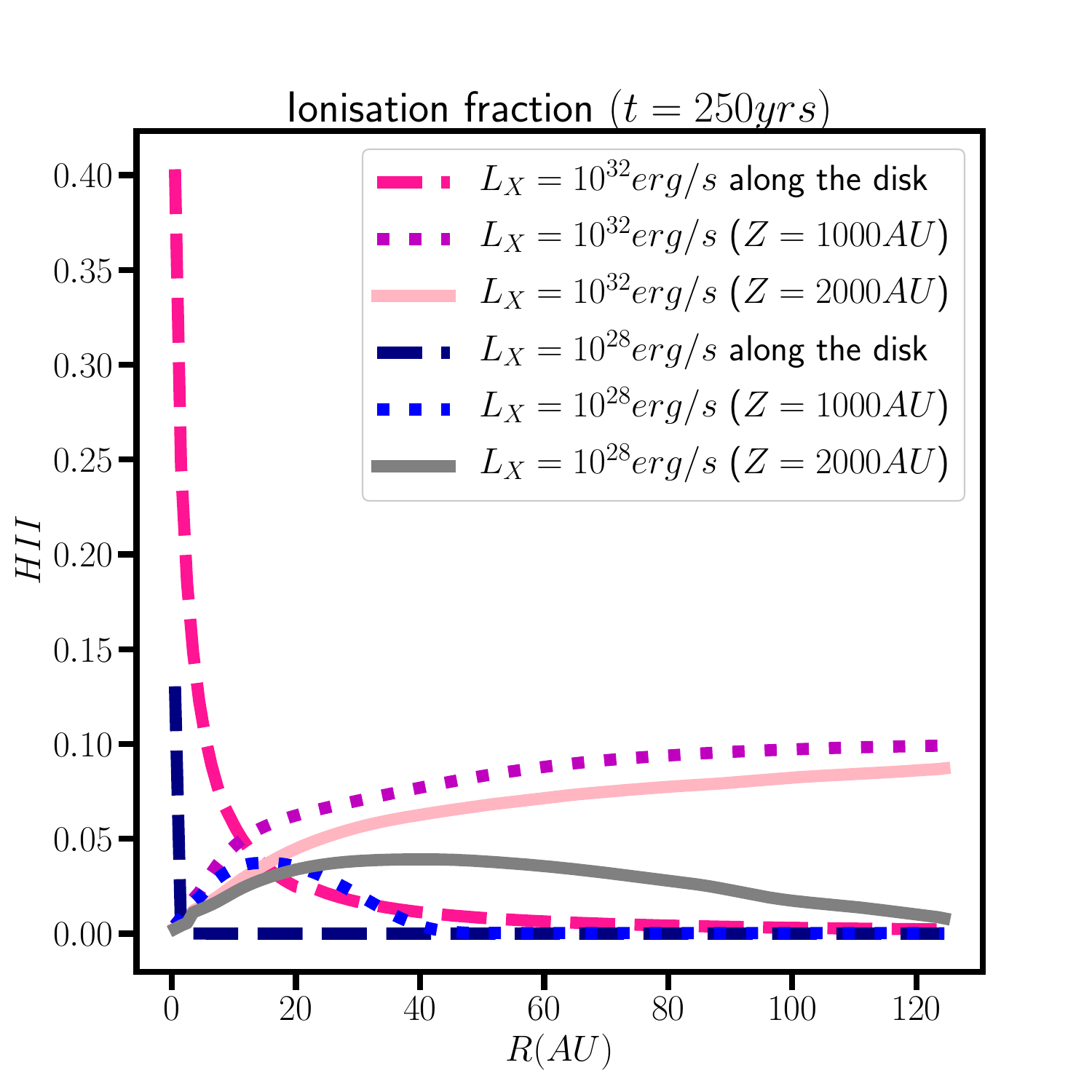}
 
  \caption {\footnotesize Ionisation fraction profiles for different luminosity $L_{\X} = 10^{28}$ erg/s,
   $L_{\X} = 10^{32}$ erg/s at different heights, corresponding to $t = 250 \, {\rm yrs}$. }
  \label{pro.eps}
\end{figure}

In the second case (Fig.~\ref{fig:ionization_fraction}, lower panels) we compare different evolutionary snapshots for two different simulations, corresponding to a high luminosity case ($L_{\X}= 10^{32}\,$ erg/s, upper half of each panel) and a low luminosity case ($L_{\X}= 10^{28}\,$ erg/s, lower half of each panel). While the latter does not produce any noticeable difference in the amount of ionization when compared to the purely radiative case (see Fig.~\ref{fig:ionization_fraction_SNEq}), the former results in an appreciably larger ionization rate. Close to the central object we now observe almost $\sim 50\,\%$ ionization degree, lowering to $\sim 10\,\%$ at large distances (see the zoomed image below). The ionization area is now also much larger, extending to $R = 120$ AU with $10\,\%$ ionization fraction and reaching the outer radial boundary at $z = 2000$ AU (see Fig.~\ref{pro.eps}). Ions can survive large-distance propagation owing to the rather long recombination timescale.

\begin{figure*}[!h]
  \includegraphics[trim=13cm 25cm 0cm 0cm,clip,width=0.9\textwidth]{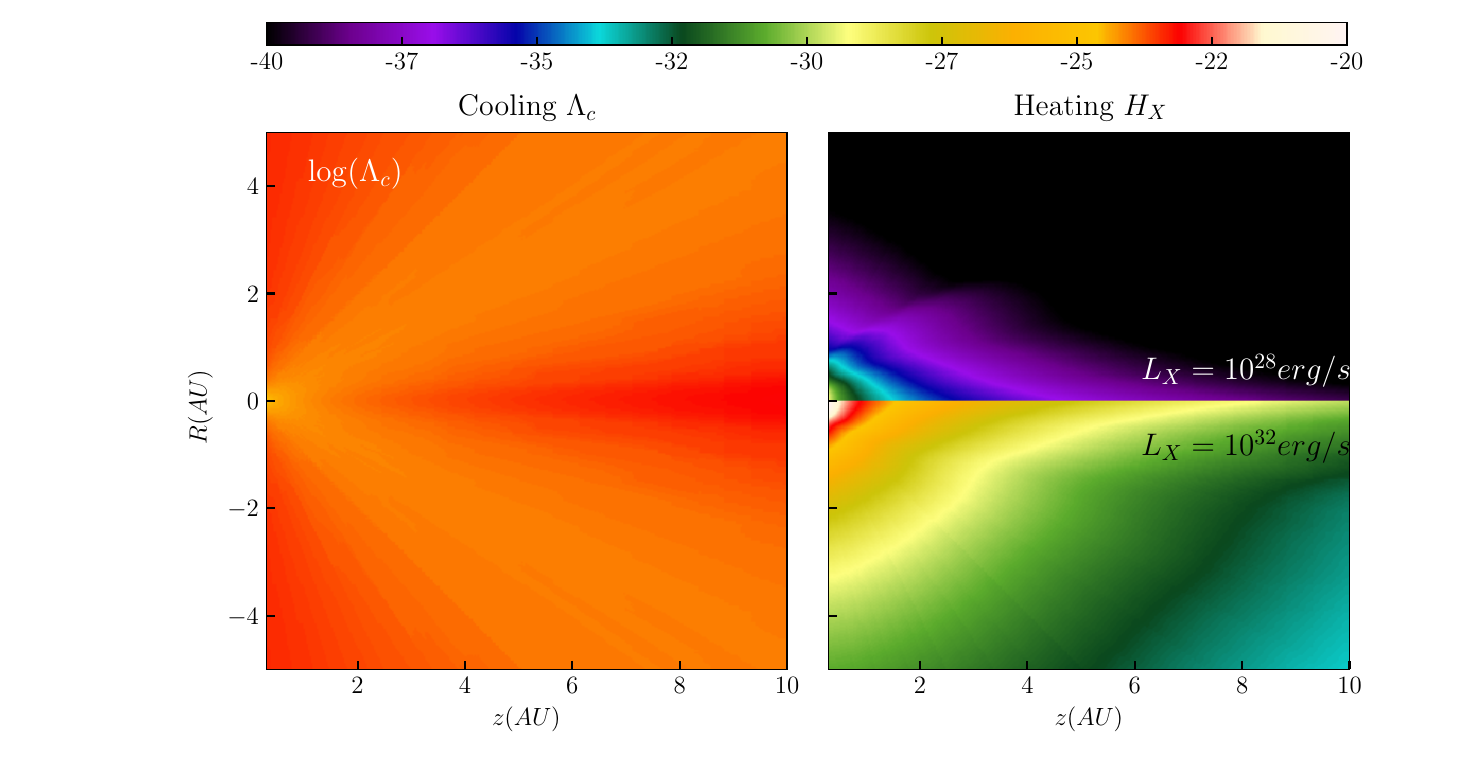}%mapfr
  \caption{Left panel: Zoomed logarithmic 2D map of the cooling $\Lambda_{c}$ in $erg \quad cm^{-3} \quad s^{-1}$. Right panel: Zoomed image of the logarithmic 2D heating map $ \mathcal{H}_\X$ in $erg \quad cm^{-3} \quad s^{-1}$, for luminosity $L_{\X} = 10^{32} erg /s$ and $L_{\X} = 10^{28} erg /s$, at $t = 150\, {\rm yrs}$ . }
  \label{fig:neutral_fraction}
\end{figure*}

A close-up of the central region is shown in the top panel of Fig.~\ref{fig:ionization_fraction} at $t = 150\, {\rm yrs}$ for both the low and high luminosity cases. Here we show a composite image of the ionization fractions for $z < 130$ AU (bottom larger panel) and a smaller view of the central region ($z < 10$ AU) (upper left panel). From the figure it is evident that the jet is ionized by $\sim 50\,\%$ close to the star ($R \lesssim 2$ AU) owing to the strong photo-ionizing X-ray flux of the high-luminosity case. This fraction decreases to the nominal value of $10$--$20\,\%$ and is essentially transported as the jet propagates outwards.

In Fig.~\ref{fig:neutral_fraction} we include both cooling and heating maps, the latter case being split into the low- and high-luminosity cases. Heating is stronger close to the origin, reaching peak values in the range $10^{-20}-10^{-23}\,{\rm erg\,cm^{-3}\,s^{-1}}$ and thus appreciably larger than cooling. Heating is then quickly suppressed with distance as the luminosity decreases as $1/R^2$, becoming negligible already beyond a distance of $\approx 3$ AU where cooling overwhelms ($\Lambda_{c}$ is about $10^{-24}\,{\rm erg\,cm^{-3}\,s^{-1}}$). This result is in agreement with the work of \cite{TE08}.

These results confirm and validate the assumption that photoionizing X-rays from the central star can be held responsible for the inferred ionization rates at very large distances \cite{TE12}. They have shown that shocks propagating along the jet in a medium that has been pre-ionized at a ratio of about $20\,\%$ give off-line emission from the post-shock region that well matches the observed surface brightness, especially at distances above a few hundred AU. Without pre-ionization, the post-shock brightness falls short by about an order of magnitude with respect to observations of the jet in RW Aurigae \cite{TE12}.

\section{Summary and conclusions}

By means of high-resolution MHD numerical computations we have investigated the effect of stellar-driven X-ray photoionization on the jet launching and its subsequent large-scale propagation up to 2000 AU. Owing to the large disparities in temporal and spatial scales, the numerical simulations were conducted using adaptive mesh refinement with the PLUTO code in 2D cylindrical axisymmetric coordinates. The outflow is magnetically driven by an accretion disk imposed as a boundary condition. Both optically thin cooling and X-ray heating have been included in our model.

Owing to the small X-ray photon mean-free-path and geometrical dilution, heating is effective only at a distance of a few AU from the central star. For typical X-ray luminosities in classical T Tauri stars ($10^{28}$ erg/s $\lesssim L_{\X} \lesssim 10^{32}$ erg/s), we find that photoionization is capable of ionizing the jet material up to $\sim 50\,\%$ (for the highest luminosity case) close to the central object. Because of the slow recombination rates, the ionized material can survive large-scale propagation, reaching steady-state asymptotic values of $\sim 10-20\,\%$ far from the launching region, in agreement with the assumption made by \cite{TE12}.

We have also confirmed that the ejection rate, temperature, and velocity lie within the typical ranges of observed astronomical sources.

In future work, we plan to model specific YSO jets with the inclusion of shocks along the beam generated by variability in the ejection process at the base of the outflow. Synthetic maps will be produced with a post-processing code that calculates the emission starting from the gas physical parameters determined by the output of the dynamical simulation. Such maps will be compared with observed maps at high angular resolution.

\textbf{Acknowledgements} We thank an anonymous referee for his/her criticisms, suggestions and comments that have greatly helped to improve the paper. We are grateful to Dr. Bhargav Vaidya, Asst. Professor, Centre of Astronomy (IIT Indore), Indian Institute of Technology Indore for insightful discussion. We acknowledge the CINECA award under the ISCRA initiative, for the availability of high performance computing resources and support. Z. Ahmane was supported by the Programme National Exceptionnel (P.N.E) within the Ministry of Higher Education of Algeria and by the predoc scholarship funded by the University of Torino.

\bibliographystyle{apalike} % Choose a style
\bibliography{pp}  

\end{document}